\documentclass[10pt, journal]{IEEEtran}
\usepackage{moreverb}
\usepackage{graphicx}
\usepackage{multirow}
\usepackage{amsmath,amsfonts}
\usepackage{amsthm}
\usepackage{pxfonts}
\usepackage{txfonts}
\usepackage{graphicx}
\usepackage{multirow}
\usepackage{mathrsfs}
\usepackage{eso-pic}
\usepackage{fancyhdr}
\usepackage{algorithm}
\usepackage{algpseudocode}
\usepackage{cite}
\usepackage{color}

\newtheorem{theorem}{Theorem}
\newtheorem{lemma}[theorem]{Lemma}
\newtheorem{proposition}[theorem]{Proposition}
\newtheorem{corollary}[theorem]{Corollary}

\newcommand{\tabincell}[2]{\begin{tabular}{@{}#1@{}}#2\end{tabular}}

\begin{document}
\title{Rooting out the Rumor Culprit from Suspects}

\author{Wenxiang Dong, Wenyi Zhang, and Chee Wei Tan

\thanks{The work of W. Dong and W. Zhang has been supported by National Basic Research Program of China (973 Program) through grant 2012CB316004, and by the 100 Talents Program of Chinese Academy of Sciences. The work of C. W. Tan has been supported by the Research Grants Council of Hong Kong under Project No. RGC CityU 125212. The material in this paper was presented in part at the IEEE International Symposium on Information Theory (ISIT), Istanbul, Turkey, July 2013.

W. Dong and W. Zhang are with Department of Electronic Engineering and Information Science, University of Science and Technology of China, Hefei 230027, China. (Email: javin@mail.ustc.edu.cn, wenyizha@ustc.edu.cn)

C. W. Tan is with Department of Computer Science, City University of Hong Kong, 83 Tat Chee Avenue, Hong Kong SAR, China. (Email: cheewtan@cityu.edu.hk)}
}

\maketitle

\begin{abstract}
  Suppose that a rumor originating from a single source among a set of suspects spreads in a network, how to root out this rumor source? With the {\it a priori} knowledge of the set of suspect nodes and a snapshot observation of infected nodes, we construct a maximum {\it a posteriori} (MAP) estimator to identify the rumor source using the susceptible-infected (SI) model. {When analyzing the performance of the MAP estimator, the {\it a priori} suspect set and its associated connectivity in the network bring about new ingredients to the problem. For this purpose, we propose to use {\it local rumor center}, which is a generalized concept based on the notion of rumor centrality, to identify the source from suspects.} For regular tree-type networks with node degree $\delta$, we characterize $\mathbf{P_c}(n)$, the correct detection probability of the source estimator upon observing $n$ infected nodes, in both the finite and asymptotic regimes. First, when the suspect set degenerates into the entirety of the network, so that every infected node belongs to the suspect set, $\lim_{n \rightarrow \infty}\mathbf{P_c}(n)$ grows from 0.25 to 0.307 as $\delta$ increases from three to infinity, a result first established in Shah and Zaman (2011, 2012) via a different approach with more mathematical machinery; furthermore, $\mathbf{P_c}(n)$ monotonically decreases with $n$ and increases with $\delta$ even in the finite-$n$ regime. Second, when the suspect nodes form a connected subgraph of the network, $\lim_{n \rightarrow \infty}\mathbf{P_c}(n)$ significantly exceeds the {\it a priori} probability if $\delta \geq 3$, and reliable detection is achieved as $\delta$ becomes sufficiently large; furthermore, $\mathbf{P_c}(n)$ monotonically decreases with $n$ and increases with $\delta$. Third, when there are only two suspect nodes, $\lim_{n \rightarrow \infty}\mathbf{P_c}(n)$ is at least 0.75 if $\delta \geq 3$; and $\mathbf{P_c}(n)$ increases with the distance between the two suspects. Fourth, when there are multiple suspect nodes, among all possible connection patterns, that all the suspects form a single connected subgraph of the network achieves {the smallest detection probability} for the MAP source estimator. Our analysis leverages ideas from the P\'{o}lya's urn model in probability theory and sheds insight into the behavior of the rumor spreading process {not only in the asymptotic regime but also for the general finite-$n$ regime}.
\end{abstract}


\section{Introduction}

Spreading of epidemics and information cascades through social networks is ubiquitous in the modern world \cite{Kempe_simp,Wasserman_sna}. Examples include the propagation of infectious diseases, information diffusion in the Internet, tweeting and retweeting of popular Twitter topics. In general, any of these situations can be modeled as an epidemic-like rumor spreading in a network \cite{Shah_rsd2011,Bailey_sir}. A key challenge is to identify the rumor source by leveraging the network topology, suspect characteristics and the observation of infected nodes. Finding this rumor source has practical applications and also allows us to better understand the role of the network in the spreading amplification effect.

In practical scenarios, there is often {\it a priori} knowledge that certain nodes are more likely to be the rumor source and the others are less likely so. Indeed, there are many situations where only a {portion} of nodes in the network have the potential to initiate the rumor spreading. When an infectious disease is spread, the frequent travellers from earlier infected cities have a higher suspicion of causing an epidemic outbreak in a new city. When a network of water pipes is polluted by microorganisms or chemical substances, only the suspicious vulnerable points need to be examined to identify the pollution source. When the cyberspace is hit by rumors or computer viruses, most of the victims are in fact innocent and thus not all victims should be suspected as the culprit. Therefore, it is not necessary to treat every infected node as the possible rumor source when identifying the source. Hereafter, we call the nodes that have the potential to initiate the rumor spreading in a network as {\it suspect nodes}. Indeed, as we will see in this paper, the suspect characteristics significantly affect detectability, and add an interesting dimension to identifying the source reliably.

In this paper, we consider the issue of reliably identifying a single rumor source from among a number of suspects, conditioned on a snapshot observation of infected nodes in a network. The rumor spreading process is modeled by the susceptible-infected (SI) model, a special case of the general susceptible-infected-recovered (SIR) model for infectious disease spreading \cite{Bailey_sir}. In the SI model, there are two types of nodes: \romannumeral1) susceptible nodes that are capable of becoming infected, \romannumeral2) infected nodes that can spread the rumor; and spreading occurs in a cascading manner, i.e., once a susceptible node gets infected by its neighbor, it remains infected (having the rumor) and in turn may infect its other susceptible neighbors. Our goal is to identify the rumor source based only on the knowledge of network structure, the set of suspect nodes and the snapshot observation of the infected nodes. The {\it a priori} suspect set and its characteristics like connectivity bring about new ingredients and challenges to the detection problem.

\subsection{Related Works}

Over the past decade, abundant works have dealt with the problems on epidemic outbreaks across networks. The main focus has been to understand the impacts of network structure and the infection/cure rates of diffusion processes, such as in \cite{Moore_epidemic,Satorras_epidemic,Newman_epidemic,Lind_epidemic,Ganesh_epidemic}. Besides, many researchers have developed network inference techniques to learn the underlying network parameters and predict the propagation characteristics, such as in \cite{Streftaris_inf,Demiris_inf,Okamura_inf,Rodriguez_inf}. In addition, the issue of extracting the influential source nodes for epidemic spreadings has also been considered, such as in \cite{Kempe_simp,Chen_simp,Dong_simp}.

The rumor source estimation problem has only been recently studied in the pioneering work \cite{Shah_rsd2011,Shah_rsd2012} for a single rumor source identification using the SI model. In particular, for regular tree-type networks, \cite{Shah_rsd2011,Shah_rsd2012} proposed a maximum likelihood (ML) estimator and derived its asymptotic performance. When the node degree $\delta=2$, i.e., for a linear network, the asymptotic correct detection probability is zero; when $\delta=3$, it is 0.25; when $\delta>3$, it is a positive constant value $\phi_1(\delta)$, which approaches $0.307$ as $\delta$ grows large. Besides, other types of networks and models have also been considered, such as geometric trees \cite{Shah_rsd2011}, random graphs \cite{Shah_rsdarXiv}, identification of multiple rumor sources in the SI model \cite{Luo_rsd} and its counterpart in the SIR model \cite{Zhu_rsd}, and noisy estimation of a single source \cite{Pinto_rsd}. However, all these works assume that every node in the network has the potential to initiate the rumor spreading.

\subsection{Our Contributions}

In this paper, we study the single rumor source estimation problem with the {\it a priori} knowledge that the rumor source is restricted to a specified {\it suspect set} $S$ of suspect nodes in a network $G$. We construct a maximum {\it a posteriori} (MAP) estimator to identify the rumor source, and focus on the analysis of $\mathbf{P_c}(n)$, the correct detection probability of the estimator upon observing $n$ infected nodes. For regular tree-type networks with node degree $\delta$, we characterize $\mathbf{P_c}(n)$ in the {\it finite} and the {\it asymptotic} (with the number of infected nodes tending large) regimes, using the P\'{o}lya's urn model in probability theory \cite{Johnson_urn} and assuming a uniform {\it a priori} distribution of the suspects in $S$. {To handle the analysis of the MAP estimator, we develop a key concept of {\it local rumor center}, which generalizes the notion of rumor centrality in \cite{Shah_rsd2011} and proves to be a handy tool considerably facilitating the analysis. We also highlight new insights for the detection in general networks with suspects.} We find that the introduction of the {\it a priori} knowledge of suspect set substantially enriches, rather than simplifies, the rumor source estimation problem.

Our main contributions are summarized as follows.

1) When $S$ contains all the nodes in $G$, our estimator reduces to that in \cite{Shah_rsd2011,Shah_rsd2012}, recovering the same asymptotic detection performance results on $\mathbf{P_c}(n)$ as theirs. Our proofs are different and appear to be more natural intuitively, requiring less mathematical machinery. Furthermore, we obtain exact detection performance results on $\mathbf{P_c}(n)$ for regular trees with any finite $n$ and $\delta$. In addition, we prove that $\mathbf{P_c}(n)$ monotonically decreases with $n$ and increases with $\delta$ even in the finite regime. This result formalizes our intuition that the earlier the observation is made the more reliable it is to estimate the rumor source.

2) In the case where $S$ with cardinality $k$ forms a connected subgraph of $G$, we obtain exact detection performance results on $\mathbf{P_c}(n)$ for regular trees in both finite and asymptotic regimes. Asymptotically, $\lim_{n \rightarrow \infty}\mathbf{P_c}(n)$ is always {a positive constant value $\phi_2(\delta,k)$}. Notably, $\phi_2(\delta,k)$ significantly exceeds the {\it a priori} probability $1/k$ for regular trees with degree larger than two, and is at least $1/2$ even with arbitrarily large cardinality $k$. Furthermore, the MAP estimator achieves reliable detection, i.e., $\lim_{n \rightarrow \infty}\mathbf{P_c}(n) \rightarrow 1$ as $\delta$ grows sufficiently large. We also prove that $\mathbf{P_c}(n)$ monotonically decreases with $n$ and increases with $\delta$ even in the finite regime. It is a striking observation that {$\mathbf{P_c}(n)$ is at least $1/2$ (with $\delta > 2$) given the {\it a priori} knowledge of suspects --- even for arbitrarily large $k$ as long as it is finite, in contrast to that $\mathbf{P_c}(n)$ is at most $1/2$ with no {\it a priori} knowledge in the case where every node in the network has the potential to initiate the rumor spreading \cite{Shah_rsd2011}.} Along with the analysis, we present an interpretation of that key observation.

3) In the case where $S$ contains only two suspect nodes separated by their shortest path distance $d$ (measured by number of hops), we obtain exact detection performance results on $\mathbf{P_c}(n)$ in the finite regime; and when $\delta \geq 3$, we show that $\lim_{n \rightarrow \infty}\mathbf{P_c}(n) \geq 0.75$, which furthermore approaches one as $\delta$ grows sufficiently large. In addition, $\mathbf{P_c}(n)$ increases with the separation distance $d$ for general regular trees.

4) When $S$ with cardinality $k$ forms a general subgraph of $G$, we prove that the MAP estimator achieves {the smallest detection probability when all the suspects form a single connected subgraph of $G$.} This result formalizes our intuition that the more clustered the suspects are the more difficult it is to identify the rumor source.

\subsection{Organization}

First, Section \uppercase\expandafter{\romannumeral2} describes the framework of the SI rumor spreading model and the MAP estimator to identify the rumor source {in general networks}. Second, in Section \uppercase\expandafter{\romannumeral3} we establish analytical results on the detection probability of the MAP estimator for regular tree-type networks, under four representative scenarios. Third, we design and implement algorithms to compute the exact detection probability of the MAP estimator for regular tree-type networks in Section \uppercase\expandafter{\romannumeral4}. Fourth, we carry out simulation experiments to corroborate and illustrate our analytical results in Section \uppercase\expandafter{\romannumeral5}. Finally we conclude the paper in Section \uppercase\expandafter{\romannumeral6}.

\section{Rumor Spreading Model and Rumor Source Estimator}

In this section, we describe the SI rumor spreading model, give the MAP estimator for the rumor source in {general networks}, and introduce the notion of local rumor center. To facilitate the reader, {we list the key parameters used in the paper in Table I.}

\begin{table}[ht]
  \centering
  \caption{{Key Terms and Symbols}}
  \begin{tabular}{|c|l|} \hline
  Symbol & Definition\\ \hline
  $\delta$ & node degree\\ \hline
  $G_n$ & snapshot observation of $n$ infected nodes\\ \hline
  $T_u^s$ & subtree rooted at node $u$ with node $s$ as the source\\ \hline
  $R(u,G_n)$ & rumor centrality of node $u$ in $G_n$\\ \hline
  $S$ & suspect set\\ \hline
  $s_i,i \in N_{+}$ & suspect node\\ \hline
  $s^\ast$ & original rumor source\\ \hline
  $\hat{s}$ & estimate of the rumor source\\ \hline
  $s$ & rumor source by assumption\\ \hline
  $\mathbf{P_s}(\cdot)$ & prior distribution of the rumor source over the suspects\\ \hline
  $\mathbf{P_G}(\cdot)$ & \tabincell{l}{probability distribution of an infection sample in the rumor\\
                                        spreading process or an infection sample equivalently\\
                                        constructed using the P\'{o}lya's urn model}\\ \hline
  $\mathbf{P_c}(\cdot)$ & correct detection probability\\ \hline
  $\mathbf{P_e}(\cdot)$ & error detection probability, i.e., the complement of $\mathbf{P_c}(\cdot)$ \\ \hline
  $\mathbf{P_{e1}}(\cdot)$ & error detection probability by one specified subtree\\ \hline
  \end{tabular}
\end{table}

\subsection{Rumor Spreading Model}

In general, an undirected network $G=(V,E)$ consists of a set of nodes $V$ and a set of edges $E$. The set of nodes $V$ is assumed countably infinite so as to avoid any boundary effect, and any pair of nodes may infect each other if and only if they are connected by an edge in $E$. The network is assumed to be static, i.e., no node joins or leaves.

We consider the case where only a single node in an {\it a priori} specified {\it suspect set} $S$ is the original rumor source who initiates the infection process among the susceptible nodes in $G$, where $S=\{s_1,s_2,\cdots,s_k\} \subseteq V$ has cardinality $k$. Nodes in $S$ are called {\it suspect nodes}. We assume an {\it a priori} distribution $\mathbf{P_s}$ of the rumor source over the nodes in $S$, and thus $\mathbf{P_s}(s)$ denotes the probability that $s \in S$ initiates the rumor spreading. For normalization, we have
\begin{equation}
\label{eq1}
\sum_{i=1}^k{\mathbf{P_s}(s_i)} = 1.
\end{equation}
Throughout this paper, we assume a uniform distribution $\mathbf{P_s}$ over the nodes in $S$, i.e., $\mathbf{P_s}(s)=1/k$ for any $s \in S$. Admittedly this uniform prior assumption is a restriction, but we remark that our analysis can be extended to handle the more general case of non-uniform priors, with slightly more technical modifications.

The rumor spreading process in an SI model unfolds as follows. Initially, only a single node $s^\ast \in S$ possesses a rumor to spread over the network $G$. Once a node has the rumor, it is termed as infected. An infected node may infect its neighbors, independent of all the other nodes. Let $\tau_{ij}$ be the time it takes for node $j$ to receive the rumor from its neighbor $i$ after $i$ has the rumor, where  $(i,j) \in E$. We assume that $\{\tau_{ij},(i,j) \in E\}$ are mutually independent and exponentially distributed with a normalized rate $\lambda = 1$ (without loss of generality).

\subsection{Rumor Source Estimator: Maximum a Posteriori (MAP)}

(1) MAP estimator

Suppose that a rumor originates from a node $s^\ast \in S$. We observe the network $G$ at some point in time and find a snapshot of $n$ infected nodes, which are collectively denoted by $G_n$. Due to the SI model, $G_n$ must form a connected subgraph of $G$ and contain at least a node in $S$, which obviously includes $s^\ast$. Our goal is to construct an estimator to identify a node $\hat{s}$ as the rumor source. The detection is correct if $\hat{s}=s^\ast$.

Now, conditioned on $G_n$, the source node has a uniform distribution over the nodes in $S \bigcap G_n$ to initiate the spreading. Utilizing the Bayes' rule, the maximum {\it a posteriori} (MAP) estimator of $s^\ast$ maximizes the {\it average} correct detection probability and is given by
\begin{eqnarray}
\label{eq2}
\hat{s} &\in& \operatorname*{arg\,max}_{s \in \{S \bigcap G_n\}}{\mathbf{P_G}\left(s|G_n\right)}\nonumber\\
&=& \operatorname*{arg\,max}_{s \in \{S \bigcap G_n\}}\frac{\mathbf{P_G}\left(G_n|s\right)\mathbf{P_s}(s)}{\mathbf{P_G}\left(G_n\right)}\nonumber\\
&=& \operatorname*{arg\,max}_{s \in \{S \bigcap G_n\}}{\mathbf{P_G}\left(G_n|s\right)},
\end{eqnarray}
where $\mathbf{P_G}(G_n|s)$ is the probability of observing $G_n$ assuming $s$ to be the rumor source. Note that a key difference from the model in \cite{Shah_rsd2011} is that in our work the rumor source resides in {\it a priori} suspect set $S \subseteq V$. Naturally, we would first evaluate $\mathbf{P_G}\left(G_n|s\right)$ for all $s \in \{S \bigcap G_n\}$ and then pick the one with the maximal value to be $\hat{s}$.

(2) Optimal MAP estimator for regular trees

In general, the evaluation of $\mathbf{P_G}\left(G_n|s\right)$ may be computationally prohibitive since it is related to counting the number of linear extensions of a partially ordered set \cite{Shah_rsd2011,Brightwell_poset}. Therefore, we leverage the concept of {\it rumor centrality}, first introduced in \cite{Shah_rsd2011}, which would enable an implementation of our estimator using only $\mathcal{O}(n)$ computation steps. In particular, the optimal MAP estimator for a regular tree is given by
\begin{equation}
\label{eq3}
\hat{s} \in \operatorname*{arg\,max}_{s \in \{S \bigcap G_n\}}{\mathbf{P_G}\left(G_n|s\right)} = \operatorname*{arg\,max}_{s \in \{S \bigcap G_n\}}{R\left(s,G_n\right)},
\end{equation}
where $R(s,G_n)$ is the {\it rumor centrality} of node $s$ in $G_n$, and can be evaluated by
\begin{equation}
\label{eq6}
R(s,G_n)=n! \prod_{u \in G_n}\frac{1}{\left|T_u^s\right|},
\end{equation}
where $T_u^s$ is the subtree rooted at node $u$ with node $s$ as the source in $G_n$ and $\left|T_u^s\right|$ is the number of nodes in $T_u^s$; e.g., see Fig. 1.

\begin{figure}[t]
\label{fig1}
\center
  \includegraphics[width=0.30\textwidth]{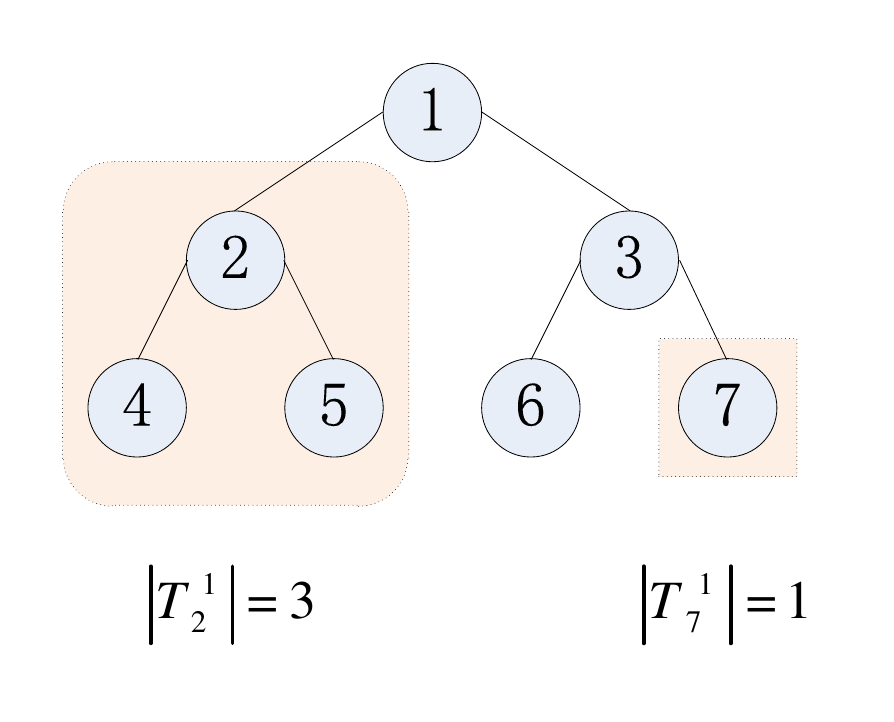}\\
  \caption{Illustration of subtree $T_u^s$.}
\end{figure}

(3) Approximate estimator for general trees and graphs

For general trees, the estimator \eqref{eq3} with the rumor centrality can be a reasonable heuristic. So, an approximate estimator is given by
\begin{equation}
\label{eq4}
\hat{s} \in \operatorname*{arg\,max}_{s \in \{S \bigcap G_n\}}{R\left(s,G_n\right)}.
\end{equation}
Besides, a message-passing algorithm has been proposed in \cite{Shah_rsd2011} to compute the rumor centralities for all the nodes in a general tree $G_n$ with $n$ nodes using only $\mathcal{O}(n)$ computation steps. For general graphs, the estimator \eqref{eq3} using the rumor centrality can be leveraged as a heuristic. Intuitively, a rumor tends to travel from the source to each infected node along a minimum-distance path \cite{Shah_rsd2011,Pinto_rsd}, and thus we approximate the diffusion tree by a breadth-first search (BFS) tree. So an approximate estimator may be given by
\begin{equation}
\label{eq5}
\hat{s} \in \operatorname*{arg\,max}_{s \in \{S \bigcap G_n\}}{R\left(s,T_{\mbox{\footnotesize bfs}}(s)\right)},
\end{equation}
where $T_{\mbox{\footnotesize bfs}}(s)$ is the BFS tree with node $s$ as its root in a general graph $G_n$. The complexity of implementing \eqref{eq5} is $\mathcal{O}(n^3)$.

\subsection{Local Rumor Center in General Trees}

{In the following, we develop a notion of local rumor center, which enables an efficient implementation of the estimator \eqref{eq3} and will be instrumental for our subsequent analysis.} {The local rumor center may be viewed as a ``conditional'' rumor center, and renders the performance analysis of the MAP estimator \eqref{eq3} tractable.} A useful recursive relationship is that, for any two neighboring nodes $u$ and $v$ in a tree $G_n$ \cite{Shah_rsd2011},
\begin{equation}
\label{eq7}
R(u,G_n)=R(v,G_n) \frac{\left|T_u^v\right|}{n-\left|T_u^v\right|}.
\end{equation}

Now, consider a node $\omega$ with a neighbor set $N(\omega)$ and a sub-neighborhood $N_l(\omega) \subseteq N(\omega)$. If $R(\omega,G_n) \geq R(u,G_n)$ for all $u \in N_l(\omega)$, then $\omega$ is called the {\it local rumor center} with respect to (w.r.t.) the sub-neighborhood $N_l(\omega)$ of $G_n$. More precisely, the {\it local rumor center} is characterized as follows.

\begin{proposition}
{\romannumeral1}) Given a tree $G_n$ of $n$ nodes, if node $\omega$ is the local rumor center w.r.t. a sub-neighborhood $N_l(\omega) \subset G_n$, then for any $u \in N_l(\omega)$, we have $\left|T_{u}^{\omega}\right| \leq {n}/{2}$; and for any $u^\prime \in T_u^{\omega} \setminus \{u\}$, we have $R(u^\prime,G_n) < R(\omega,G_n)$.\\
{\romannumeral2}) If there is a node $\omega$ such that $\left|T_u^{\omega}\right| \leq {n}/{2}$ for all $u \in N_l(\omega)$, then $\omega$ is a local rumor center w.r.t. the sub-neighborhood $N_l(\omega) \subset G_n$.\\
{\romannumeral3}) Furthermore, if node $\omega$ is the local rumor center w.r.t. a sub-neighborhood $N_l(\omega) \subset G_n$, then there is at most a node $u \in N_l(\omega)$ such that $R(u,G_n) = R(\omega,G_n)$, which holds if and only if $\left|T_u^{\omega}\right| = {n}/{2}$.
\end{proposition}

\emph{Remark 1:} In fact, the local rumor center is a generalization of the {\it rumor center} in \cite{Shah_rsd2011}, which is defined as the node with the maximal rumor centrality in $G_n$. When $N_l(\omega) = N(\omega)$, the two notions coincide. However, the rumor center may belong to the set $G_n \setminus S$ and thus is not the solution of the estimator \eqref{eq3}. Besides, note that we can find at most two local rumor centers from a connected suspect set $S$ w.r.t. the sub-neighborhood restricted by $S$, when we are given $G_n$. {The notion of local rumor center not only generalizes the concept of rumor center, but also will prove to be a key to tackle the MAP estimator with suspects.}

\begin{proof}[Proof of Proposition 1]
First consider $u \in N_l(\omega)$. From \eqref{eq7}, we have
\begin{equation}
\label{eq13}
\frac{R(u,G_n)}{R(\omega,G_n)}=\frac{\left|T_u^{\omega}\right|}{n-\left|T_u^{\omega}\right|}.
\end{equation}
Since ${R(u,G_n)} \leq {R(\omega,G_n)}$, we get $\left|T_u^{\omega}\right| \leq n/2$.

Next, consider $u^\prime \in T_u^{\omega} \setminus \{u\}$, and let $\mathcal{P}(u,u^\prime)$ be the set of nodes along the shortest path from $u$ to $u^\prime$, but not including $u$. Repeatedly using \eqref{eq7}, we have
\begin{equation}
\label{eq14}
\frac{R(u^\prime,G_n)}{R(u,G_n)}=\prod_{v \in \mathcal{P}(u,u^\prime)}\frac{\left|T_v^{\omega}\right|}{n-\left|T_v^{\omega}\right|}.
\end{equation}
Since $\left|T_u^{\omega}\right| \leq n/2$, thus $\left|T_v^{\omega}\right| < \left|T_u^{\omega}\right| \leq n/2$, i.e. $\left|T_v^{\omega}\right|/(n-\left|T_v^{\omega}\right|) < 1$ for all $v \in \mathcal{P}(u,u^\prime)$. Therefore, $R(u^\prime,G_n) < R(u,G_n) \leq R(\omega,G_n)$. This proves Proposition 1-{\romannumeral1}.

Now for Proposition 1-{\romannumeral2}, if $\left|T_u^{\omega}\right| \leq {n}/{2}$, i.e. ${\left|T_u^{\omega}\right|}/{(n-\left|T_u^{\omega}\right|)} \leq 1$ for all $u \in N_l(\omega)$, then from \eqref{eq13}, we have ${R(u,G_n)} \leq {R(\omega,G_n)}$. Therefore, $\omega$ is the local rumor center w.r.t. its sub-neighborhood $N_l(\omega) \subset G_n$.

As for Proposition 1-{\romannumeral3}, if there is a node $u \in N_l(\omega)$ such that $R(u,G_n) = R(\omega,G_n)$, then from \eqref{eq13}, we have $\left|T_u^{\omega}\right| = {n}/{2}$; the deduction holds backward. Besides, there can be at most a subtree $T_u^{\omega}$ ($u \in N_l(\omega)$) such that $\left|T_u^{\omega}\right| = n/2$, since the total number of nodes in $G_n$ is $n$ and any two subtrees with $\omega$ as the source are disjoint. As a result, if node $\omega$ is the local rumor center w.r.t. a sub-neighborhood $N_l(\omega) \subset G_n$ and there has already been a node $u^\ast \in N_l(\omega)$ such that $R(u^\ast,G_n) = R(\omega,G_n)$, then for all $u \in N_l(\omega)\setminus\{u^\ast\}$ we have $\left|T_u^{\omega}\right| < n/2$, i.e., $\left|T_u^{\omega}\right|/(n-\left|T_u^{\omega}\right|) < 1$. Again, from \eqref{eq13}, we have ${R(u,G_n)} < {R(\omega,G_n)}$.
\end{proof}

\section{Detection Probability in Regular Trees: Analytical Results}

In this section, we analyze the performance of the MAP estimator for regular tree-type networks. Four representative scenarios are investigated: when the suspect set contains all the nodes in the network, i.e., with no {\it a priori} knowledge; when the suspect set forms {a connected subgraph} of the network; when the suspect set contains only two nodes; when the suspect set contains multiple nodes which are possibly disconnected. In analysis we exploit the key fact that the rumor spreading process on regular trees is equivalent to the ball drawing process in the P\'{o}lya's urn model \cite{Johnson_urn}, and thus establish the performance results of the MAP estimator in both finite and asymptotically regimes. In the following, we introduce our main results and provide their proofs.

\subsection{Main Results}

We focus on the correct detection probability $\mathbf{P_c}(n)$, the probability of the MAP estimator correctly identifying the rumor source from the suspect set $S$ upon observing $G_n$ of $n$ infected nodes in the network $G=(V,E)$. For a regular tree with node degree $\delta$, we have the following four characterizations for $\mathbf{P_c}(n)$.

(1) The case where all nodes are suspects

\begin{figure}[t]
\label{fig2}
\center
  \includegraphics[width=0.375\textwidth]{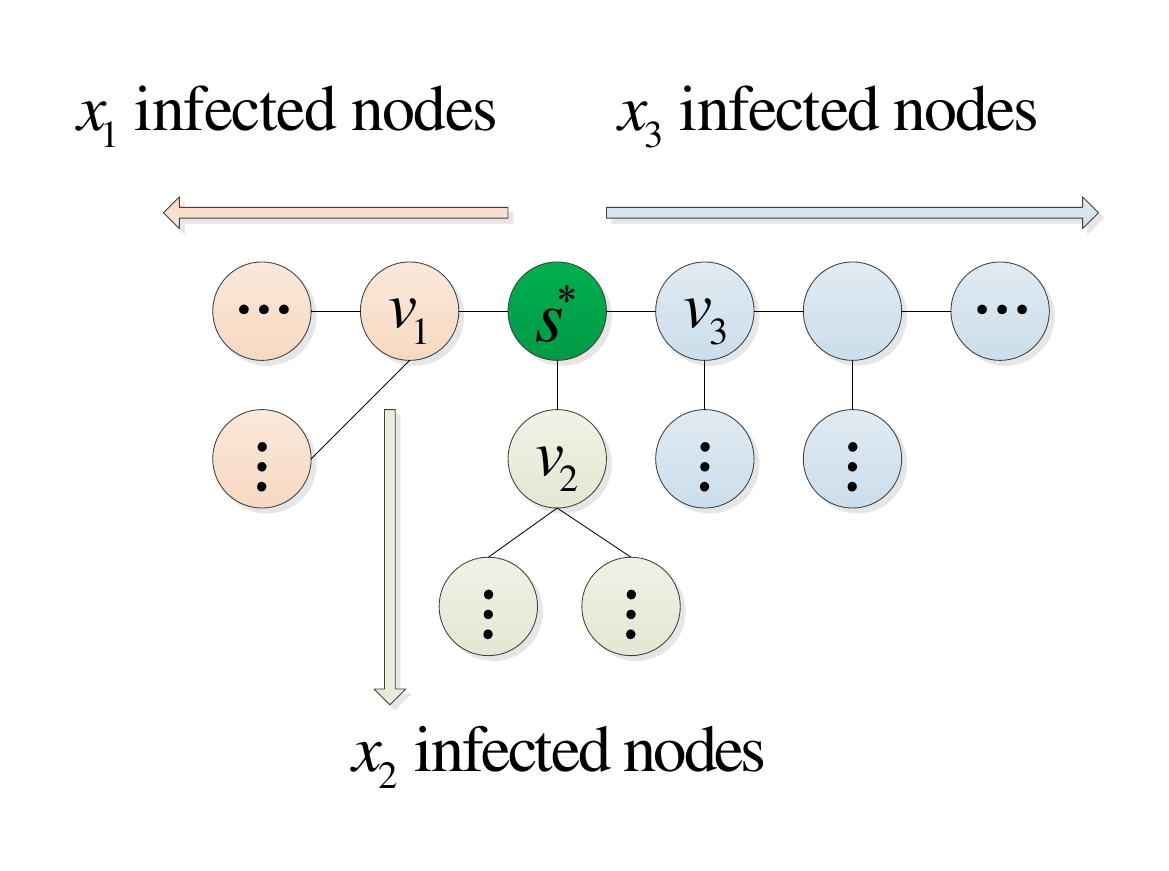}\\
  \caption{Illustration of $X_1=x_1$, $X_2=x_2$, $X_3=x_3$ infected nodes in subtrees $T_{v_1}^{s^\ast}$, $T_{v_2}^{s^\ast}$ and $T_{v_3}^{s^\ast}$ with $s^\ast$ as the source on a regular tree $G$ with node degree $\delta=3$, respectively.}
\end{figure}

In this case, $S=V$, every infected node in an observation $G_n$ has the potential to be the rumor source, i.e., without any {\it a priori} knowledge. Assuming a node $s^\ast$ to be the rumor source with $\delta$ neighbors $v_1,\ldots,v_\delta$, we could observe $X_j=x_j$ ($1 \leq j \leq \delta$) infected nodes in each subtree $T_{v_j}^{s^\ast}$ rooted at $v_j$ with $s^\ast$ as the source in $G_n$, where $X_j$ is a random variable; e.g., see Fig. 2. Then, we will establish the following theorem.

\begin{theorem}
Suppose $S=V$, i.e., every infected node is a suspect node, then:\\
{\romannumeral1}) When $\delta=2$ (linear network),
\begin{equation}
\label{eq21}
\mathbf{P_c}(n)=\frac{1}{2^{n-1}} \binom{n-1}{\lfloor(n-1)/2\rfloor},
\end{equation}
and $\mathbf{P_c}(n)=\mathcal{O}(1/\sqrt{n})$ with sufficiently large $n$.\\
{\romannumeral2}) When $\delta=3$,
\begin{equation}
\label{eq22}
\mathbf{P_c}(n)=\frac{1}{4}+\frac{3}{4} \frac{1}{2\lfloor n/2 \rfloor +1},
\end{equation}
and $\mathbf{P_c}(n)=0.25+\mathcal{O}(1/n)$ with sufficiently large $n$.\\
{\romannumeral3}) When $\delta>3$, we use Algorithm 1 in Section IV to compute the exact value of $\mathbf{P_c}(n)$ with finite $n$. Besides,
\begin{equation}
\label{eq23}
\lim_{n \rightarrow \infty}
\mathbf{P_c}(n) = \phi_1(\delta) := 1-\delta\left(1-I_{1/2}\left(\frac{1}{\delta-2},\frac{\delta-1}{\delta-2}\right)\right),
\end{equation}
where $I_x(\alpha,\beta)$ is the incomplete Beta function with parameters $\alpha$ and $\beta$; and $\phi_1(\delta) \rightarrow 1-\ln{2} \approx 0.307$ as $\delta \rightarrow \infty$.
\end{theorem}

\emph{Remark 2:} When $S = V$, the MAP estimator \eqref{eq3} in effect reduces to the ML estimator established in \cite{Shah_rsd2011,Shah_rsd2012}.\footnotemark[1] The asymptotic parts of Theorem 2 have been established in \cite{Shah_rsd2011,Shah_rsd2012} using a different approach from ours that explicitly relies on the exponential distribution of the infection time. Therefore, without any {\it a priori} knowledge, the estimator achieves a strictly positive detection probability if the network is not linear, but it is asymptotically upper bounded by 0.307.

\footnotetext [1] {Strictly speaking, our setup of a uniform {\it a priori} distribution of the rumor source over $S$ is not well defined when $S = V$, since $V$ is a countably infinite set. Our remedy is that we consider ML estimation for Theorem 2 thus returning to the setup in \cite{Shah_rsd2011,Shah_rsd2012}, and consider MAP estimation for the other two cases.}

(2) The case of connected suspects

\begin{figure}[t]
\label{fig3}
\center
  \includegraphics[width=0.5\textwidth]{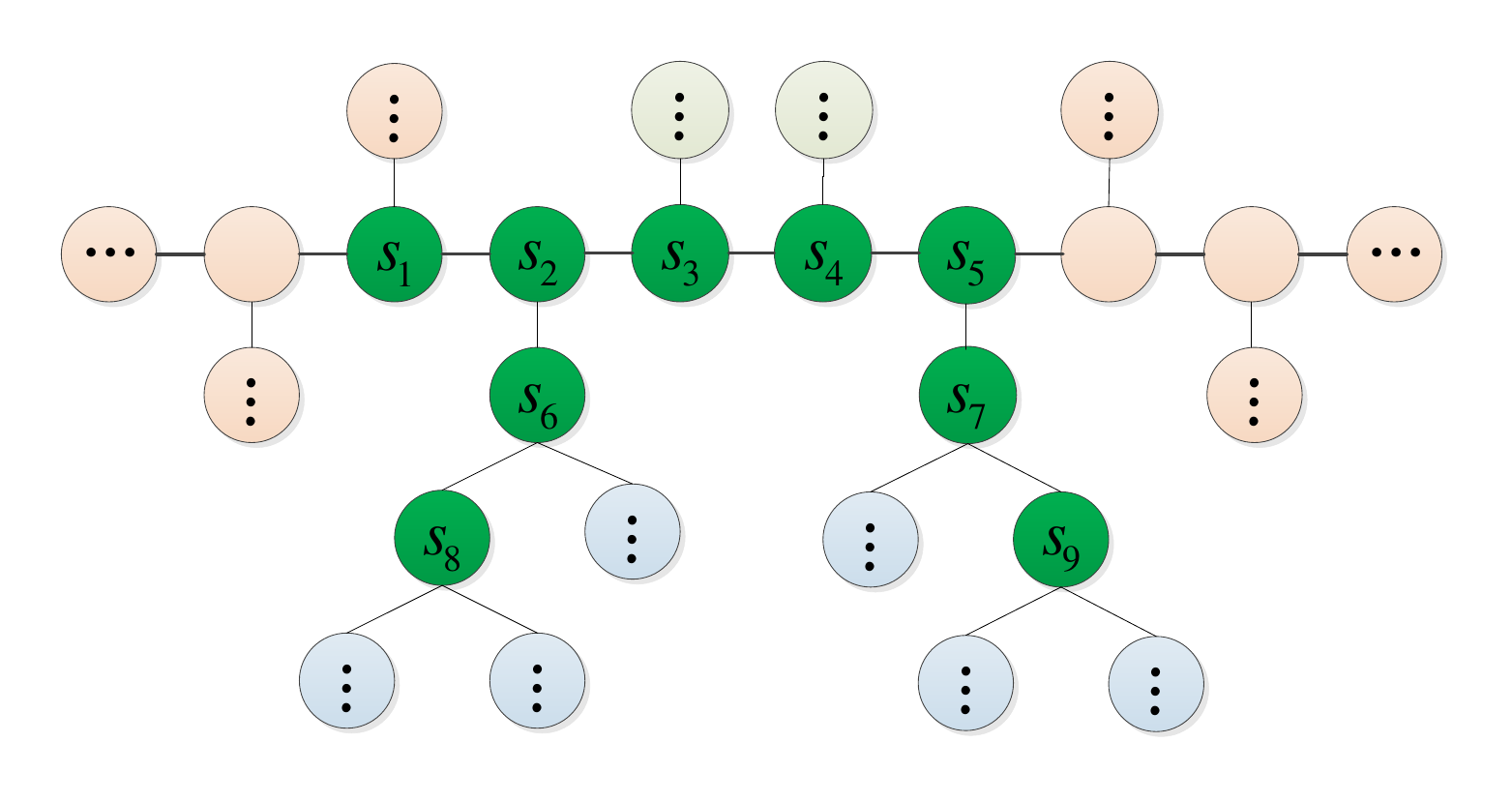}\\
  \caption{Illustration of a suspect set $S=\{s_1,s_2,\ldots,s_9\}$ with multiple connected suspect nodes on a regular tree $G$ with node degree $\delta=3$. The nodes in $S$ form a connected subgraph of $G$.}
\end{figure}

In this case, $S=\{s_1,\ldots,s_k\}$ of cardinality $k$ forms a connected subgraph of $G$; e.g., see Fig. 3. With this {\it a priori} knowledge, we will establish the following theorem.

\begin{theorem}
{Suppose that $S$ forms a connected subgraph of $G$, then:}\\
{\romannumeral1}) When $\delta=2$ (linear network),
\begin{equation}
\label{eq36}
\mathbf{P_c}(n)=\frac{1}{k}\left(1 + \frac{k-1}{2^{n-1}} \binom{n-1}{\lfloor(n-1)/2\rfloor}\right),
\end{equation}
and $\mathbf{P_c}(n)={1}/{k}+\mathcal{O}(1/\sqrt{n})$ with sufficiently large $n$.\\
{\romannumeral2}) When $\delta=3$,
\begin{equation}
\label{eq37}
\mathbf{P_c}(n)=\frac{k+1}{2k} + \frac{k-1}{k} \frac{1}{4\lfloor n/2 \rfloor +2},
\end{equation}
and $\mathbf{P_c}(n)=(k+1)/(2k)+\mathcal{O}(1/n)$ with sufficiently large $n$.\\
{\romannumeral3}) When $\delta>3$, we use Algorithm 2 in Section IV to compute the exact value of $\mathbf{P_c}(n)$ with finite $n$. Besides,
\small
\begin{equation}
\label{eq38}
\lim_{n \rightarrow \infty}
\mathbf{P_c}(n)= \phi_2(\delta,k) := 1-\frac{2k-2}{k}\left(1-I_{1/2}\left(\frac{1}{\delta-2},\frac{\delta-1}{\delta-2}\right)\right),
\end{equation}
\normalsize
where $\phi_2(\delta,k) \rightarrow 1$ as $\delta \rightarrow \infty$, and $\phi_2(\delta,k) \rightarrow 2I_{1/2}\left(\frac{1}{\delta-2},\frac{\delta-1}{\delta-2}\right)-1$ as $k \rightarrow \infty$.
\end{theorem}

\emph{Remark 3:} For linear networks, $\mathbf{P_c}(n)$ can barely exceed the {\it a priori} probability $1/k$. When $\delta \geq 3$, {$\mathbf{P_c}(n)$ is at least $\max\{1/k, 1/2\}$}, which is in sharp contrast to that $\mathbf{P_c}(n)$ is at most $1/2$ with no {\it a priori} knowledge as in Theorem 2. Furthermore, the MAP estimator achieves reliable detection as $\delta$ grows sufficiently large. Therefore, the performance of the MAP estimator is significantly improved and reliable detection can be achieved when the {\it a priori} knowledge on the suspect nodes is given.

Remarkably, even with $k \rightarrow \infty$ our results do not degenerate to the case of no {\it a priori} knowledge. This is a seemingly counter-intuitive fact, because one would expect that as $k$ grows large Theorem 3 would recover the results in Theorem 2 (and thus \cite{Shah_rsd2011}). To understand this fact, the key is that, no matter how large $k$ is, the suspect set has a boundary in the infinite regular tree network. Those suspect nodes nearby the boundary are easier to identify than those further inside the suspect set. Since the performance of the MAP estimator is the correct detection probability {\it averaged} among all suspect nodes, asymptotically reliable estimation can be attained, and such a phenomenon is especially pronounced when the node degree $\delta$ is large as then most of the suspect nodes are located nearby the boundary. In particular, we need to note that this scaling behavior does not rely on $n$; that is, $k$ and $n$ can separately grow large, without dependency between each other, --- indeed $k$ can grow even faster than $n$. We therefore believe that Theorem 3 reveals a fundamental distinction between the MAP and the ML estimation philosophies, and indicates that the introduction of the {\it a priori} knowledge of suspect set does enrich, rather than simplify, the rumor source estimation problem.

(3) The case of two suspects

\begin{figure}[t]
\label{fig4}
\center
  \includegraphics[width=0.5\textwidth]{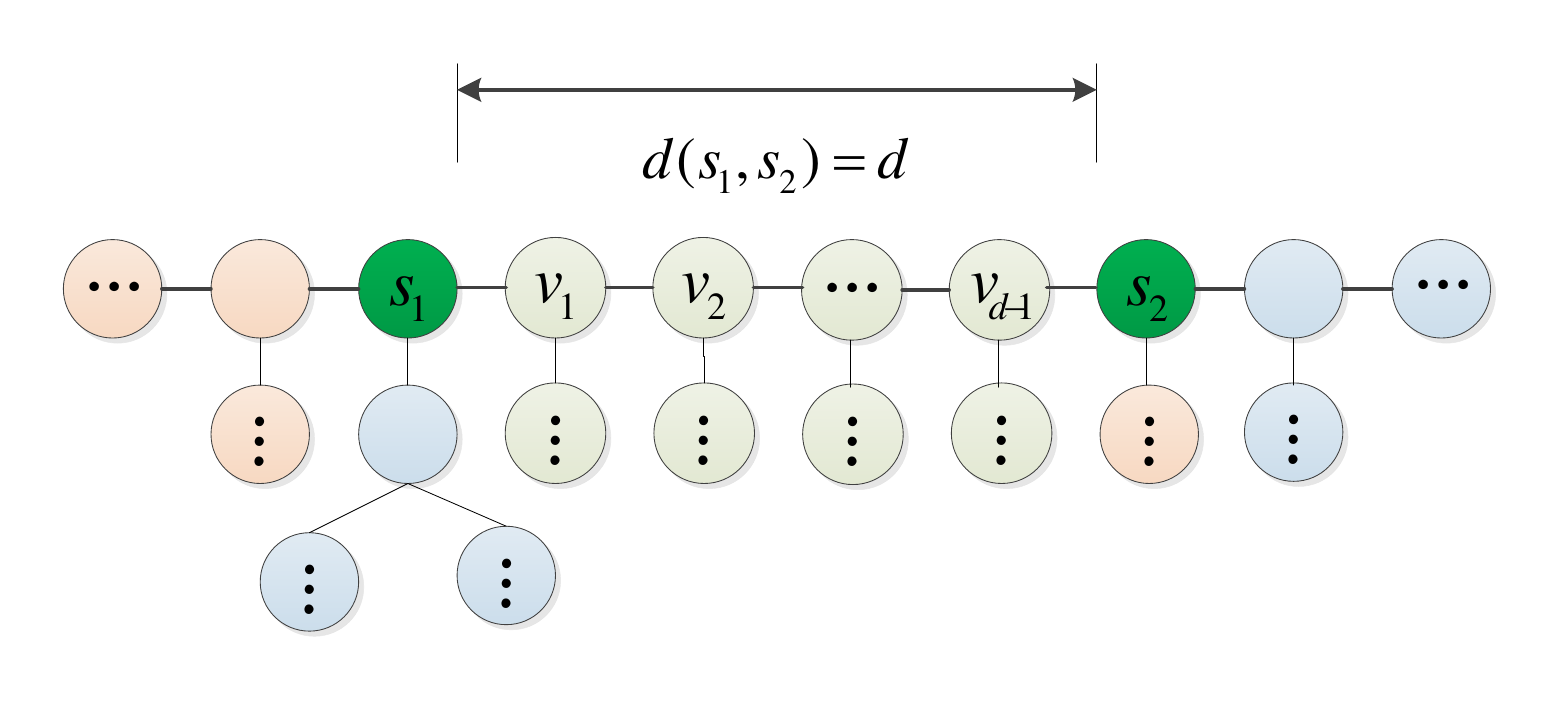}\\
  \caption{Illustration of two suspect nodes with distance $d$ on a regular tree $G$ with node degree $\delta=3$. There are $d$ nodes from $s_1$ to $s_2$ excluding $s_1$, which are sequentially denoted by $v_1,v_2,\ldots,v_{d-1},v_d=s_2$.}
\end{figure}

In this case, $S$ contains only two suspect nodes $s_1$ and $s_2$. We let $d$ be their shortest path distance on $G$ (i.e., the number of hops from $s_1$ to $s_2$); e.g., see Fig. 4. Since we can obviously ensure correct estimation of the source if $d \geq n$, we assume $d<n$. With this {\it a priori} knowledge of $S$, we have the following theorem.

\begin{theorem}
Suppose $S$ only contains two suspect nodes, and denote by $d$ their shortest path distance ($d<n$), then:\\
{\romannumeral1}) When $\delta=2$ (linear network),
\begin{equation}
\label{eq57}
\mathbf{P_c}(n)=
\begin{cases}
\begin{aligned}
\frac{1}{2}-\frac{1}{2^n}\sum_{z_1=(n-d-1)/2}^{(n+d+1)/2}{\binom{n-1}{z_1}}, \,\, (n-d) \, \mbox{is odd};
\end{aligned}\\
\begin{aligned}
\frac{1}{2}-\frac{1}{2^n}\sum_{z_1=(n-d)/2}^{(n+d-2)/2}{\binom{n-1}{z_1}}, \,\, (n-d) \, \mbox{is even}.
\end{aligned}
\end{cases}
\end{equation}
{\romannumeral2}) When $\delta=3$, we use Algorithm 3 in Section IV to compute the exact value of $\mathbf{P_c}(n)$ with finite $n$. Besides,
\begin{equation}
\label{eq58}
\lim_{n \rightarrow \infty}\mathbf{P_c}(n)
\begin{cases}
\begin{aligned}
= 0.75, \,\, d=1,
\end{aligned}\\
\begin{aligned}
\approx 0.886, \,\, d=2.
\end{aligned}
\end{cases}
\end{equation}
{\romannumeral3}) When $\delta>3$, we use Algorithm 3 in Section IV to compute the exact value of $\mathbf{P_c}(n)$ with finite $n$. Besides,
\begin{equation}
\label{eq59}
\lim_{n \rightarrow \infty}\mathbf{P_c}(n) = \phi_3(\delta) := I_{1/2}\left(\frac{1}{\delta-2},\frac{\delta-1}{\delta-2}\right), \, d=1,
\end{equation}
and $\phi_3(\delta)\rightarrow 1$ as $\delta\rightarrow \infty$.\\
{\romannumeral4}) In general, $\mathbf{P_c}(n)$ monotonically increases with $d$.
\end{theorem}

\emph{Remark 4:} Theorem 4 formalizes our intuition that it is more difficult to correctly identify the rumor source if the two suspect nodes are closer. We see that when $\delta \geq 3$, the {\it a priori} probability $1/2$ can be significantly exceeded.

(4) The case of multiple suspects with general connectivity

\begin{figure}[ht]
\label{fig5}
\center
  \includegraphics[width=0.5\textwidth]{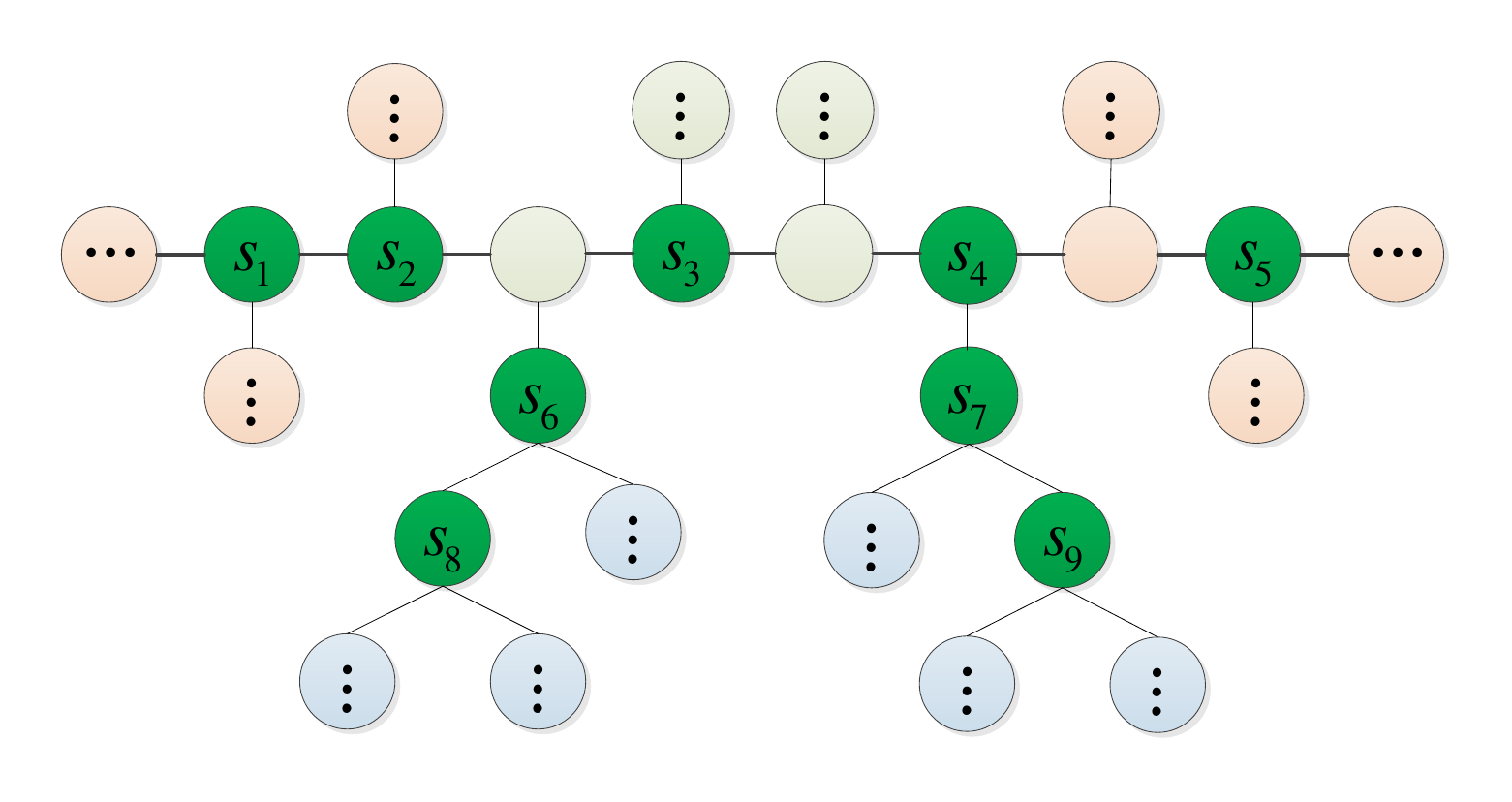}\\
  \caption{Illustration of a suspect set $S=\{s_1,s_2,\ldots,s_9\}$ with multiple suspect nodes on a regular tree $G$ with node degree $\delta=3$. The nodes in $S$ form a general subgraph of $G$.}
\end{figure}

In this case, $S=\{s_1,\ldots,s_k\}$ of cardinality $k$ forms a general subgraph of $G$; e.g., see Fig. 5. Note that this subgraph can be disconnected in general. With the {\it a priori} knowledge of $S$, we will establish the following theorem.

\begin{theorem}
{Suppose $S$ contains $k$ suspect nodes, then:}\\
{\romannumeral1}) When $\delta=2$ (linear network),
\begin{equation}
\label{eq72}
\mathbf{P_c}(n) \geq \frac{1}{k}\left(1 + \frac{k-1}{2^{n-1}} \binom{n-1}{\lfloor(n-1)/2\rfloor}\right),
\end{equation}
and $\mathbf{P_c}(n) \geq {1}/{k}+\mathcal{O}(1/\sqrt{n})$ with sufficiently large $n$.\\
{\romannumeral2}) When $\delta=3$,
\begin{equation}
\label{eq73}
\mathbf{P_c}(n) \geq \frac{k+1}{2k} + \frac{k-1}{k} \frac{1}{4\lfloor n/2 \rfloor +2},
\end{equation}
and $\mathbf{P_c}(n) \geq (k+1)/(2k)+\mathcal{O}(1/n)$ with sufficiently large $n$.\\
{\romannumeral3}) When $\delta>3$,
\small
\begin{equation}
\label{eq74}
\lim_{n \rightarrow \infty}
\mathbf{P_c}(n) \geq \phi_2(\delta,k) := 1-\frac{2k-2}{k}\left(1-I_{1/2}\left(\frac{1}{\delta-2},\frac{\delta-1}{\delta-2}\right)\right).
\end{equation}
{\romannumeral4}) \normalsize In general, $\mathbf{P_c}(n)$ is minimized, among all possible $S$'s with $|S| = k$, when the $k$ suspect nodes constitute a connected subgraph in $G$ as in Theorem 3.
\end{theorem}

\emph{Remark 5:} Theorem 5 can be established as a consequence of Theorems 3 and 4. When there are $k$ suspect nodes, {the MAP estimator achieves the smallest detection probability in the scenario of Theorem 3.} This result formalizes our intuition that the more clustered the suspects are the more difficult it is to identify the rumor source.

\subsection{Equivalence to the P\'{o}lya's Urn Model}

(1) Preliminaries

We first introduce the preliminaries on the P\'{o}lya's urn model in probability theory, and then show that the rumor spreading process on regular trees is equivalent to the ball drawing process in the P\'{o}lya's urn model, a connection first utilized in \cite{Shah_rsd2012} for asymptotic analysis. Our approach is to leverage it to derive exact detection probability in the more general finite regime (recovering the asymptotic result as a special case).

\normalsize
P\'{o}lya's urn model \cite[Chap. 4]{Johnson_urn}: initially, the urn contains $b_j$ balls of color $C_j$ ($1 \leq j \leq \delta$); at each uniform drawing of a single ball, the ball is returned together with $\epsilon$ balls of the same color; after $n$ drawings, the number $X_j$ is the number of times that the balls of color $C_j$ are drawn. Then the joint distribution of $\{X_j,1 \leq j \leq \delta\}$ is given by

\small
\begin{equation}
\label{eq15}
{\mathbf{P_G}}\left[\bigcap_{j=1}^{\delta}(X_j=x_j)\right]=
\frac{n!}{x_1! x_2! \cdots x_\delta!}\frac{\prod_{j=1}^{\delta}b_j(b_j+\epsilon)\cdots(b_j+(x_j-1)\epsilon)}{b(b+\epsilon)\cdots(b+(n-1)\epsilon)},
\end{equation}

\normalsize
\noindent where $b=\sum_{j=1}^{\delta}b_j$ and $\sum_{j=1}^{\delta}x_j=n$.

{
As $n \rightarrow \infty$, the limiting joint distribution of the ratios $\{{X_j}/{n},1 \leq j \leq \delta\}$ converges to the Dirichlet distribution, whose density function is given by
\begin{equation}
\label{eq16}
\frac{\Gamma(\alpha)}{\prod_{j=1}^{\delta}\Gamma(\alpha_j)} \prod_{j=1}^{\delta}y_j^{\alpha_j-1},
\end{equation}
where $\alpha_j=b_j/\epsilon$, $\alpha=\sum_{j=1}^{\delta}\alpha_j$ and $\sum_{j=1}^{\delta}y_j=1$. Here, $\Gamma(\alpha)$ is the Gamma function with parameter $\alpha$.
}

Besides, the marginal distribution of $X_1$ is
\begin{equation}
\label{eq17}
{\mathbf{P_G}}\left(X_1=x_1\right)=
\frac{n!}{x_1^\prime! x_2^\prime!}\frac{\prod_{j=1}^{2}b_j^\prime(b_j^\prime+\epsilon)\cdots(b_j^\prime+(x_j^\prime-1)\epsilon)}{b(b+\epsilon)\cdots(b+(n-1)\epsilon)},
\end{equation}
where $b_1^\prime=b_1$, $b_2^\prime=b-b_1$, $x_1^\prime=x_1$ and $x_2^\prime=n-x_1$. In fact, it can be seen as a special case of the P\'{o}lya's urn model with two colors.

{
As $n \rightarrow \infty$, the limiting marginal distribution of the ratio ${X_1}/{n}$ converges to the Beta distribution, whose density function is given by
\begin{equation}
\label{eq18}
\frac{\Gamma(\alpha_1^\prime+\alpha_2^\prime)}{\Gamma(\alpha_1^\prime)\Gamma(\alpha_2^\prime)}y_1^{\alpha_1^\prime-1} (1-y_1)^{\alpha_2^\prime-1},
\end{equation}
where $\alpha_1^\prime=\alpha_1$ and $\alpha_2^\prime=\alpha-\alpha_1$. In particular, we have
\begin{eqnarray}
\label{eq19}
&&\lim_{n \rightarrow \infty}{\mathbf{P_G}}\left(\frac{X_1}{n} \leq x\right) = I_x(\alpha_1^\prime,\alpha_2^\prime)\nonumber\\
&:=& \frac{\Gamma(\alpha_1^\prime+\alpha_2^\prime)}{\Gamma(\alpha_1^\prime)\Gamma(\alpha_2^\prime)} \int_0^x{y^{\alpha_1^\prime-1} (1-y)^{\alpha_2^\prime-1}}dy,
\end{eqnarray}
for all $x \in [0,1]$. Note that $I_x(\alpha_1^\prime,\alpha_2^\prime)$ is called the incomplete Beta function with parameters $\alpha_1^\prime$ and $\alpha_2^\prime$.
}

\begin{figure}[t]
\label{fig6}
\center
  \includegraphics[width=0.5\textwidth]{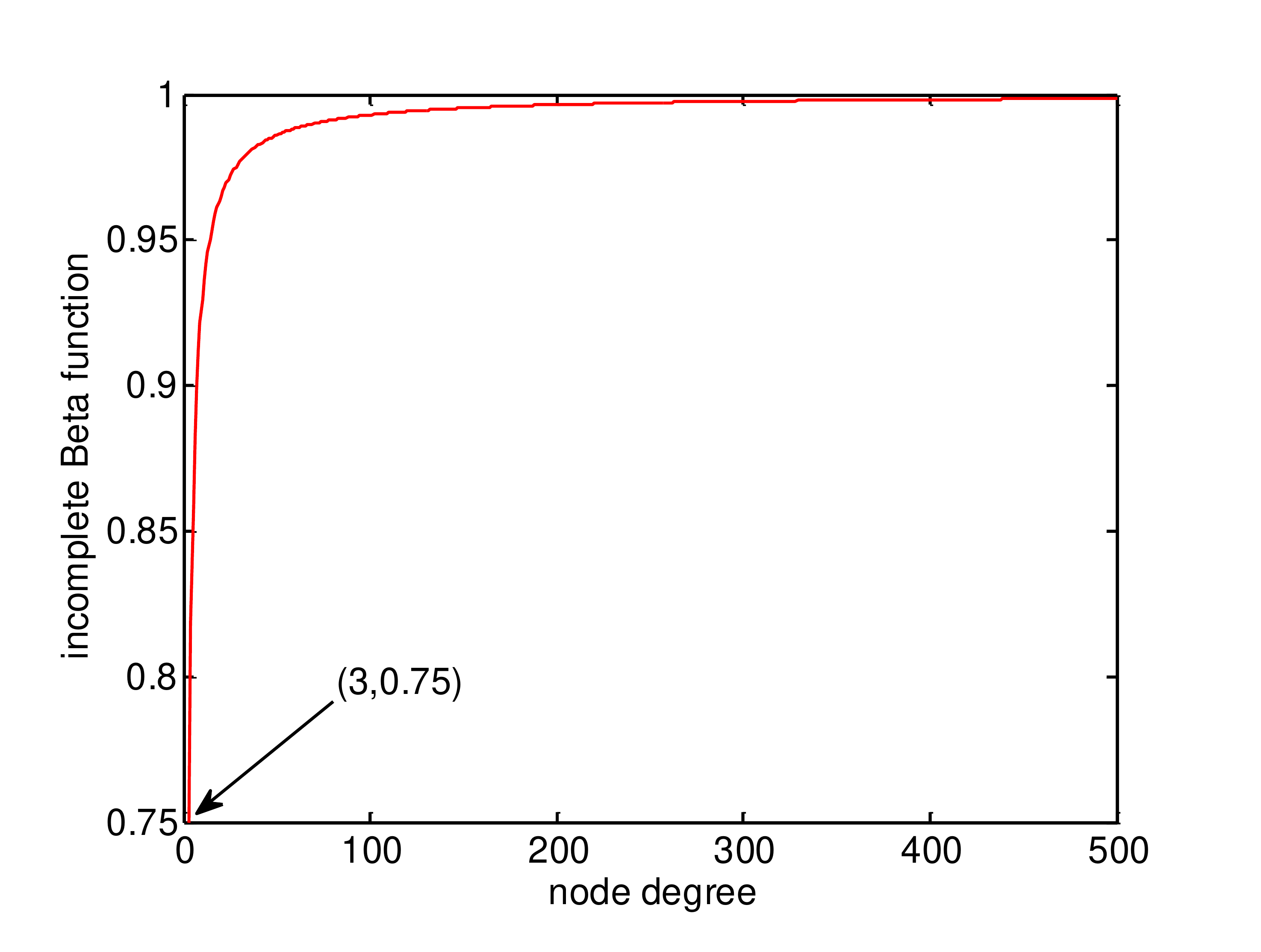}\\
  \caption{The incomplete Beta function $I_{1/2}(\frac{1}{\delta-2},\frac{\delta-1}{\delta-2})$ vs. node degree $\delta$.}
\end{figure}

We are interested in $I_{1/2}(\alpha_1^\prime,\alpha_2^\prime)$ with parameters $\alpha_1^\prime=1/(\delta-2)$ and $\alpha_2^\prime=(\delta-1)/(\delta-2)$, where $\delta$ is the node degree of a regular tree and $\delta \geq 3$; e.g., see Fig. 6.

(2) Equivalence to the P\'{o}lya's urn model

Next, we show that the rumor spreading on regular trees can be modeled by ball drawing in the P\'{o}lya's urn model, whose known distributions are used to obtain Theorems 2-5.

For a rumor source $s^\ast$ with $\delta$ neighboring nodes $v_1,\ldots,v_\delta$, let $T_{v_j}^{s^\ast}$ ($1 \leq j \leq \delta$) be the subtree rooted at node $v_j$ with node $s^\ast$ as the source in $G_n$, and define a random variable $X_j$ as the number of nodes in $T_{v_j}^{s^\ast}$; e.g., see Fig. 2. We denote the set of susceptible neighbors of infected nodes as the {\it rumor boundary}. In the rumor spreading process, nodes in $G_n$ are infected sequentially, and thus we have the following: initially, $s^\ast$ has one neighbor in each subtree $T_{v_j}^{s^\ast}$ ($1 \leq j \leq \delta$) that belongs to the rumor boundary; after one of those nodes is infected, it introduces $\delta -1$ new nodes into the rumor boundary; finally, $n-1$ nodes are infected besides $s^\ast$. Due to the memoryless property of exponential distribution and the independent and identically distributed (i.i.d.) property of the infection times $\{\tau_{ij}, (i, j) \in E\}$, in each step, the infected node is uniformly selected from the rumor boundary.

Now, the resulting infection $G_n$ with $X_j$ nodes in $T_{v_j}^{s^\ast}$ ($1 \leq j \leq \delta$) can be constructed in an equivalent way by the P\'{o}lya's urn model \cite[Chap. 4]{Johnson_urn}: initially, the urn has one ball for each color $C_j$; at each uniform drawing of a single ball, the ball is returned together with $\epsilon=\delta-2$ additional balls of the same color; after $n-1$ draws, $X_j$ is the number of times that the balls of color $C_j$ are drawn.

Therefore, in the rumor spreading process, if we assume $s^\ast$ to be the rumor source with $\delta$ neighbors $v_1,\ldots,v_\delta$ and observe $n$ infected nodes $G_n$ with $X_j$ nodes in each subtree $T_{v_j}^{s^\ast}$ ($1 \leq j \leq \delta$), then from \eqref{eq15}, the joint distribution of $\{X_j,1 \leq j \leq \delta\}$ is given by

\small
\begin{equation}
\label{eq24}
{\mathbf{P_G}}\left[\bigcap_{j=1}^{\delta}(X_j=x_j)\right]=
\frac{(n-1)!}{x_1! x_2! \cdots x_\delta!}\frac{\prod_{j=1}^{\delta}1(1+\epsilon)\cdots(1+(x_j-1)\epsilon)}{\delta(\delta+\epsilon)\cdots(\delta+(n-2)\epsilon)},
\end{equation}

\normalsize
\noindent where $\sum_{j=1}^{\delta}x_j=n-1$.

Besides, the marginal distribution of $X_1$ is
\begin{equation}
\label{eq240}
{\mathbf{P_G}}\left(X_1=x_1\right)=
\binom{n-1}{x_1}\frac{\prod_{j=1}^{2}b_j^\prime(b_j^\prime+\epsilon)\cdots(b_j^\prime+(x_j^\prime-1)\epsilon)}{\delta(\delta+\epsilon)\cdots(\delta+(n-2)\epsilon)},
\end{equation}
where $b_1^\prime=1$, $b_2^\prime=\delta-1$, $x_1^\prime=x_1$ and $x_2^\prime=n-x_1-1$.

{
We are also interested in the limiting marginal distribution of the ratio ${X_1}/{n}$ as $n \rightarrow \infty$. From \eqref{eq19}, we have
\begin{equation}
\label{eq241}
\lim_{n \rightarrow \infty}
{\mathbf{P_G}}\left(\frac{X_1}{n} \leq x\right)=\frac{\Gamma(\alpha+\beta)}{\Gamma(\alpha)\Gamma(\beta)}\int_0^x{y^{\alpha-1} (1-y)^{\beta-1}}dy
=I_x(\alpha,\beta),
\end{equation}
where $\alpha=1/(\delta-2)$, $\beta=(\delta-1)/(\delta-2)$. In other words, we have
\begin{equation}
\label{eq242}
{\mathbf{P_G}}\left(\frac{X_1}{n} \leq x\right)=I_x(\alpha,\beta)+\xi_(n,\delta),
\end{equation}
where $\lim_{n \rightarrow \infty}{\xi_(n,\delta)}=0$ for any given $\delta$.
}

(3) Markov concatenation of the P\'{o}lya's urn model

Next, we focus on the case of two suspects. Notably, in Theorem 4, we use the Markov concatenation of the P\'{o}lya's urn model to derive the probability distribution of the infection observed in the rumor spreading process, focusing on the nodes along the path between two suspects $s_1$ and $s_2$.

Assume $s_1$ to be the rumor source $s^\ast$, and let $\mathcal{P}=\{v_0=s_1,v_1,\cdots,v_d=s_2\}$ be the shortest path from $s_1$ to $s_2$; e.g., see Fig. 4. Let a random variable $Z_h$ ($1 \leq h \leq d$) be the number of nodes in subtree $T_{v_h}^{s^\ast}$ rooted at node $v_h$ with node $s^\ast$ as the source in $G_n$. It is clear that $Z_{h} \geq Z_{h+1}+1$ if $Z_{h}>0$ for all $1 \leq h \leq d-1$. In the proof of Theorem 4, we focus on the error detection probability $\mathbf{P_e}(n)=1-\mathbf{P_c}(n)$. Therefore, without loss of generality, we assume $Z_h>0$ for all $1 \leq h \leq d$. Here, we can also construct the random variables $Z_h$ ($1 \leq h \leq d$) equivalently with the concatenation use of the P\'{o}lya's urn model.

For random variable $Z_1$, we construct it using the P\'{o}lya's urn model as follows: initially, the urn has $b_1^1=1$ black ball and $b_1^1=\delta-1$ white balls; at each uniform drawing of a single ball, the ball is returned together with $\epsilon=\delta-2$ additional balls of the same color; after $n-1$ draws, the number $Z_1$ is the number of times that balls of black color are drawn. Therefore, from \eqref{eq15}, the distribution of $Z_1$ is
\begin{equation}
\label{eq60}
{\mathbf{P_G}}\left[Z_1=z_1\right]=
\binom{n-1}{z_1} \frac{\prod_{j=1}^{2}b_j^1(b_j^1+\epsilon)\cdots(b_j^1+(x_j^1-1)\epsilon)}{b^1(b^1+\epsilon)\cdots(b^1+(n-1)\epsilon)},
\end{equation}
where $b^1=\delta$, $x_1^1=z_1$ and $x_2^1=n-z_1-1$.

For random variable $Z_h$ ($2 \leq h \leq d$) conditioned on $Z_{h-1}=z_{h-1}$, we construct it using the P\'{o}lya's urn model as follows: initially, the urn has $b_b^h=1$ black ball and $b_w^h=\delta-2$ white balls; at each uniform drawing of a single ball, the ball is returned together with $\epsilon=\delta-2$ additional balls of the same color; after $z_{h-1}-1$ draws, the number $Z_h$ is the number of times that balls of black color are drawn. Again, from \eqref{eq15}, the distribution of $Z_h$ is

\small
\begin{equation}
\label{eq61}
{\mathbf{P_G}}\left[Z_h=z_h|Z_{h-1}=z_{h-1}\right]=
\binom{z_{h-1}-1}{z_h} \frac{\prod_{j=1}^{2}b_j^h\cdots(b_j^h+(x_j^h-1)\epsilon)}{b^h\cdots(b^h+(z_{h-1}-2)\epsilon)},
\end{equation}

\normalsize
\noindent where $b^h=\delta-1$, $x_1^h=z_h$ and $x_2^h=z_{h-1}-z_h-1$.

Since $Z_h$ only depends on $Z_{h-1}$ for all $2 \leq h \leq d$, thus $Z_1,Z_2,\ldots,Z_d$ form a Markov chain. Therefore, the joint distribution of $\{Z_h,1 \leq h \leq d\}$ is

\small
\begin{equation}
\label{eq62}
{\mathbf{P_G}}\left[\bigcap_{h=1}^{d}(Z_h=z_h)\right]=
{\mathbf{P_G}}\left[Z_1=z_1\right] \prod_{h=2}^{d}{\mathbf{P_G}}\left[Z_h=z_h|Z_{h-1}=z_{h-1}\right].
\end{equation}

\normalsize

\subsection{Proof of Theorem 2: Suspecting all Nodes}

In the case of $S=V$, we only need to consider an arbitrary node $s^\ast \in G$ as the rumor source by symmetry. For a source $s^\ast$ with $m$ ($m \leq \delta$) neighbors $N_l(s^\ast)=\{v_1,\ldots,v_m\} \subset S$, let a random variable $X_j$ be the number of nodes in each subtree $T_{v_j}^{s^\ast}$ ($1 \leq j \leq m$) of $G_n$. Then, we have the following lemma for the argument of Theorem 2 ($m=\delta$) and Theorem 3 ($m \leq \delta$); see its proof in Appendix-A.

\begin{lemma}
To correctly identify source $s^\ast$ with $m$ neighboring suspect nodes as the estimate $\hat{s}$, we have
\begin{equation}
\label{eq40}
\begin{cases}
\begin{aligned}
p_{1}:={\mathbf{P_c}}\left(\hat{s} = s^\ast \bigg| \max\{x_j,1 \leq j \leq m\}<n/2\right)=1,
\end{aligned}\\
\begin{aligned}
p_{1/2}:={\mathbf{P_c}}\left(\hat{s} = s^\ast \bigg| \max\{x_j,1 \leq j \leq m\}=n/2\right)=\frac{1}{2},
\end{aligned}\\
\begin{aligned}
p_{0}:={\mathbf{P_c}}\left(\hat{s} = s^\ast \bigg| \max\{x_j,1 \leq j \leq m\}>n/2\right)=0.
\end{aligned}
\end{cases}
\end{equation}
\end{lemma}

\emph{Remark 6:} Lemma 6 is deduced from Proposition 1. In order to prove Theorem 2 (and Theorem 3), we should find the conditions, under which $s^\ast$ is the local rumor center w.r.t. $N_l(s^\ast)$ of $G_n$, such that the estimator \eqref{eq3} can correctly identify $s^\ast$ as the source.

Since $S=V$, the source $s^\ast$ has $m=\delta$ neighboring suspect nodes. Using Lemma 6, we can write $\mathbf{P_c}(n)$ as
\begin{eqnarray}
\label{eq26}
\mathbf{P_c}(n)&=&p_{1/2}\cdot\sum_{\max\{x_j,1 \leq j \leq \delta\}=n/2}{\mathbf{P_G}}\left[\bigcap_{j=1}^{\delta}(X_j=x_j)\right]\nonumber\\
&&+p_{1}\cdot\sum_{\max\{x_j,1 \leq j \leq \delta\}<n/2}{\mathbf{P_G}}\left[\bigcap_{j=1}^{\delta}(X_j=x_j)\right],
\end{eqnarray}
where ${\mathbf{P_G}}\left[\bigcap_{j=1}^{\delta}(X_j=x_j)\right]$ is given by \eqref{eq24}.

In the following, we will establish the argument of Theorem 2 for three special cases when the node degree $\delta=2$, $\delta=3$ and $\delta>3$, respectively.

(1) Detection probability when $\delta=2$

\begin{proof}[Proof of Theorem 2-{\romannumeral1}]
When $\delta=2$, the distribution in \eqref{eq24} can be written as
\begin{equation}
\label{eq27}
{\mathbf{P_G}}\left[\bigcap_{j=1}^{2}(X_j=x_j)\right]=\frac{(n-1)!}{x_1! x_2!} \frac{1}{2^{n-1}}.
\end{equation}

From \eqref{eq26}, the correct detection probability is
\begin{eqnarray}
\label{eq28}
\mathbf{P_c}(n)&=&\frac{1}{2} \sum_{\max\{x_1,x_2\}=n/2}{\mathbf{P_G}}\left[\bigcap_{j=1}^{2}(X_j=x_j)\right]\nonumber\\
&&+ \sum_{\max\{x_1,x_2\}<n/2}{\mathbf{P_G}}\left[\bigcap_{j=1}^{2}(X_j=x_j)\right],\nonumber\\
&=&\frac{1}{2^{n-1}} \binom{n-1}{\lfloor(n-1)/2\rfloor}.
\end{eqnarray}
In the above, the detailed deduction is in Appendix-B.

As $n \rightarrow \infty$, by the Stirling's formula, we have
\begin{eqnarray}
\label{eq29}
\mathbf{P_c}(n)&\approx&\frac{1}{2^n} \cdot \frac{n!}{\left[(n/2)!\right]^2}\nonumber\\
&\approx&\frac{1}{2^n} \cdot \frac{\sqrt{2 \pi n} \cdot \left(\frac{n}{e}\right)^n}{\left[\sqrt{\pi n} \cdot \left(\frac{n}{2e}\right)^{n/2}\right]^2}\nonumber\\
&=&\sqrt{\frac{2}{\pi n}}\nonumber\\
&=&\mathcal{O}\left(\frac{1}{\sqrt{n}}\right).
\end{eqnarray}
\end{proof}

(2) Detection probability when $\delta=3$

\begin{proof}[Proof of Theorem 2-{\romannumeral2}]
When $\delta=3$, the distribution in \eqref{eq24} can be written as
\begin{equation}
\label{eq30}
{\mathbf{P_G}}\left[\bigcap_{j=1}^{3}(X_j=x_j)\right]=\frac{2}{n(n+1)}.
\end{equation}

From \eqref{eq26}, the correct detection probability is
\begin{eqnarray}
\label{eq31}
\mathbf{P_c}(n)&=&\frac{1}{2} \sum_{\max\{x_j,1 \leq j \leq 3\}=n/2}{\mathbf{P_G}}\left[\bigcap_{j=1}^{3}(X_j=x_j)\right]\nonumber\\
&&+ \sum_{\max\{x_j,1 \leq j \leq 3\}<n/2}{\mathbf{P_G}}\left[\bigcap_{j=1}^{3}(X_j=x_j)\right]\nonumber\\
&=&\frac{1}{4}+\frac{3}{4} \frac{1}{2\lfloor n/2 \rfloor +1}.
\end{eqnarray}
In the above, the detailed deduction is in Appendix-C.
\end{proof}

(3) Detection probability when $\delta > 3$

When $n$ is finite, we can numerically compute the exact detection probability using Algorithm 1 in Section IV. When $n \rightarrow \infty$, we can obtain the asymptotic correct detection probability similarly to those in \cite{Shah_rsd2012}.

Before the argument of Theorem 2-{\romannumeral3}, we present the following lemma, which will be also used in the argument of Theorem 3-{\romannumeral3}; see its proof in Appendix-D. For a source $s^\ast$ with $m$ ($m \leq \delta$) neighbors $N_l(s^\ast)=\{v_1,\ldots,v_m\}$ in $S$, let a random variable $X_j$ be the number of nodes in each subtree $T_{v_j}^{s^\ast}$ ($1 \leq j \leq m$) of $G_n$.

\begin{lemma}
Define $E_j=\{X_j < {n}/{2}\}$ and $F_j=\{X_j \leq {n}/{2}\}$, $1 \leq j \leq m$. To correctly identify source $s^\ast$ with $m$ neighboring suspect nodes, we have
\begin{equation}
\label{eq49}
\begin{cases}
\begin{aligned}
\mathbf{P_c}(n|s^\ast) \geq 1-m\mathbf{P_G}(E_1^c)
\end{aligned}\\
\begin{aligned}
\mathbf{P_c}(n|s^\ast) \leq 1-m\mathbf{P_G}(F_1^c)
\end{aligned}
\end{cases}
\end{equation}
where $\mathbf{P_G}(E_1^c)$ and $\mathbf{P_G}(F_1^c)$ are the probabilities that the complements of events $E_1$ and $F_1$ occur, respectively.
\end{lemma}

\emph{Remark 7:} Lemma 7 is deduced from Proposition 1, and is a generalization of the statement claimed in \cite[Section 4.1.2]{Shah_rsd2012}. In order to prove Theorem 2-{\romannumeral3} (and Theorem 3-{\romannumeral3}), we should show that the lower and upper bounds asymptotically coincide.

\begin{proof}[Proof of Theorem 2-{\romannumeral3}]
{Since $E_1=\{X_1 < {n}/{2}\}$ and $F_1=\{X_1 \leq {n}/{2}\}$, i.e., $E_1=\{X_1/n < 1/2\}$ and $F_1=\{X_1/n \leq 1/2\}$, from \eqref{eq241}, we have}
\begin{eqnarray}
\label{eq34}
\lim_{n \rightarrow \infty}\mathbf{P_G}(E_1)&=&\lim_{n \rightarrow \infty}\mathbf{P_G}(F_1)\nonumber\\
&=&\int_{y=0}^{1/2}{\frac{\Gamma(\alpha+\beta)}{\Gamma(\alpha)\Gamma(\beta)}y^{\alpha-1} (1-y)^{\beta-1}}dy\nonumber\\
&=&I_{1/2}\left(\frac{1}{\delta-2},\frac{\delta-1}{\delta-2}\right).
\end{eqnarray}

In other words, from \eqref{eq242}, we have
\begin{equation}
\label{eq341}
\mathbf{P_G}(E_1)=\mathbf{P_G}(F_1)=I_{1/2}\left(\frac{1}{\delta-2},\frac{\delta-1}{\delta-2}\right)+\xi(n,\delta),
\end{equation}
where $\lim_{n \rightarrow \infty}\xi(n,\delta)=0$.

Since $S=V$, the source $s^\ast$ has $m=\delta$ neighboring suspect nodes. Using Lemma 7, the asymptotic correct detection probability is
{
\begin{equation}
\label{eq35}
\lim_{n \rightarrow \infty} \mathbf{P_c}(n)=1-\delta\left(1-I_{1/2}\left(\frac{1}{\delta-2},\frac{\delta-1}{\delta-2}\right)\right).
\end{equation}
}

Besides, we see that the asymptotic value $\lim_{n \rightarrow \infty}{\mathbf{P_c}(n)} > 0.25$ if the node degree $\delta > 3$. Furthermore, it approaches $1-{\ln}2 \approx 0.307$ as $\delta \rightarrow \infty$.
\end{proof}

\subsection{Proof of Theorem 3: Connected Suspects}

Consider the case where $S=\{s_1,s_2,\ldots,s_k\}$ with cardinality $k$ forms a connected subgraph of the network $G$. {Here the restriction of the suspect set poses challenges for analyzing the problem, and we thus leverage the concept of {\it local rumor center} and exploit the graph structure so as to tackle the MAP source estimator.} By the Bayes' rule and the {\it a priori} knowledge that $\mathbf{P_s}(s^\ast)=1/k$ for any $s^\ast \in S$, we have
\begin{equation}
\label{eq39}
\mathbf{P_c}(n)=\sum_{i=1}^{k}\mathbf{P_s}(s_i)\mathbf{P_c}(n|s_i)=\frac{1}{k}\sum_{s^\ast \in S}\mathbf{P_c}(n|s^\ast).
\end{equation}

We first find the correct detection probability $\mathbf{P_c}(n|s^\ast)$ for each suspect node $s^\ast \in S$. Assume that $s^\ast \in S$ is the rumor source and it has $m$ ($m \leq \delta$) neighbors $N_l(s^\ast)=\{v_1,\ldots,v_m\} \subset S$. Let a random variable $X_j$ be the number of nodes in each subtree $T_{v_j}^{s^\ast}$ ($1 \leq j \leq m$) of $G_n$, then by Lemma 6, we should find the conditions that $s^\ast$ is the local rumor center w.r.t. $N_l(s^\ast)$ of $G_n$, and the estimator \eqref{eq3} can correctly identify $s^\ast$ as the source.

Using Lemma 6, we can write $\mathbf{P_c}(n|s^\ast)$ as
\begin{eqnarray}
\label{eq41}
\mathbf{P_c}(n|s^\ast)&=&p_{1/2}\cdot\sum_{\max\{x_j,1 \leq j \leq m\}=n/2}{\mathbf{P_G}}\left[\bigcap_{j=1}^{\delta}(X_j=x_j)\right]\nonumber\\
&&+p_{1}\cdot\sum_{\max\{x_j,1 \leq j \leq m\}<n/2}{\mathbf{P_G}}\left[\bigcap_{j=1}^{\delta}(X_j=x_j)\right],
\end{eqnarray}
where ${\mathbf{P_G}}\left[\bigcap_{j=1}^{\delta}(X_j=x_j)\right]$ is given by \eqref{eq24}.

In the following, we will establish the argument of Theorem 3 for three special cases when the node degree $\delta=2$, $\delta=3$ and $\delta>3$, respectively.

(1) Detection probability when $\delta=2$
\begin{proof}[Proof of Theorem 3-{\romannumeral1}]
When $\delta=2$, from \eqref{eq27}, we have
\begin{equation}
\label{eq42}
{\mathbf{P_G}}\left[\bigcap_{j=1}^{2}(X_j=x_j)\right]=\frac{(n-1)!}{x_1! x_2!} \frac{1}{2^{n-1}}.
\end{equation}

From \eqref{eq41}, the correct detection probability for a suspect node $s^\ast$ with $m$ neighbors in the suspect set $S$ is
\begin{eqnarray}
\label{eq43}
\mathbf{P_c}(n|s^\ast)&=&\frac{1}{2} \sum_{\max\{x_j,1 \leq j \leq m\}=n/2}{\mathbf{P_G}}\left[\bigcap_{j=1}^{2}(X_j=x_j)\right]\nonumber\\
&&+ \sum_{\max\{x_j,1 \leq j \leq m\}<n/2}{\mathbf{P_G}}\left[\bigcap_{j=1}^{2}(X_j=x_j)\right]\nonumber\\
&=&
\begin{cases}
\begin{aligned}
\frac{1}{2} + \frac{1}{2^n} \binom{n-1}{\lfloor(n-1)/2\rfloor},m=1;
\end{aligned}\\
\begin{aligned}
\frac{1}{2^{n-1}} \binom{n-1}{\lfloor(n-1)/2\rfloor},m=2.
\end{aligned}
\end{cases}
\end{eqnarray}
In the above, the detailed deduction is in Appendix-E.

For a suspect set $S$ with cardinality $k$ that forms a connected subgraph of a linear network $G$, we know that only the two suspect nodes at the endpoints of the sub-linear network have one neighboring suspect node, and all the other suspect nodes have two neighboring suspect nodes. Therefore, from \eqref{eq39}, we have
\begin{equation}
\label{eq44}
\mathbf{P_c}(n)=\frac{1}{k}\left[1 + \frac{k-1}{2^{n-1}} \binom{n-1}{\lfloor(n-1)/2\rfloor}\right].
\end{equation}

As $n \rightarrow \infty$, by the Stirling's formula, we have
\begin{equation}
\label{eq45}
\mathbf{P_c}(n)=\frac{1}{k}+\frac{k-1}{k}\mathcal{O}\left(\frac{1}{\sqrt{n}}\right)=\frac{1}{k}+\mathcal{O}\left(\frac{1}{\sqrt{n}}\right).
\end{equation}
\end{proof}

(2) Detection probability when $\delta=3$

\begin{proof}[Proof of Theorem 3-{\romannumeral2}]
When $\delta=3$, from \eqref{eq30}, we have
\begin{equation}
\label{eq46}
{\mathbf{P_G}}\left[\bigcap_{j=1}^{3}(X_j=x_j)\right]=\frac{2}{n(n+1)}.
\end{equation}

From \eqref{eq41}, the correct detection probability for a suspect node $s^\ast$ with $m$ neighbors in the suspect set $S$ is
\begin{eqnarray}
\label{eq47}
\mathbf{P_c}(n|s^\ast)&=&\frac{1}{2} \sum_{\max\{x_j,1 \leq j \leq m\}=n/2}{\mathbf{P_G}}\left[\bigcap_{j=1}^{3}(X_j=x_j)\right]\nonumber\\
&&+ \sum_{\max\{x_j,1 \leq j \leq m\}<n/2}{\mathbf{P_G}}\left[\bigcap_{j=1}^{3}(X_j=x_j)\right]\nonumber\\
&=&
\begin{cases}
\begin{aligned}
\frac{3}{4}+\frac{1}{4} \frac{1}{2\lfloor n/2 \rfloor +1},\,\,m=1;
\end{aligned}\\
\begin{aligned}
\frac{1}{2}+\frac{1}{2} \frac{1}{2\lfloor n/2 \rfloor +1},\,\,m=2;
\end{aligned}\\
\begin{aligned}
\frac{1}{4}+\frac{3}{4} \frac{1}{2\lfloor n/2 \rfloor +1},\,\,m=3.
\end{aligned}
\end{cases}
\end{eqnarray}
In the above, the detailed deduction is in Appendix-F.

For a suspect set $S$ with cardinality $k$ that forms a connected subgraph of a regular tree $G$ with node degree $\delta=3$, for each suspect node $s^\ast \in S$, we first find the number of its neighboring suspect nodes. Note that given $s^\ast$ with $m$ neighboring suspect nodes, $\mathbf{P_c}(n|s^\ast)$ is one subtracted by a same factor, $1/4-1/(8\lfloor n/2 \rfloor +4)$, $m$ times, each of which accounting for one neighboring suspect node of $s^\ast$ connected by an edge. Since there are $k-1$ edges connecting the $k$ suspect nodes in $S$, each edge will account for a reduction of the factor twice. Therefore, from \eqref{eq39}, we have
\begin{eqnarray}
\label{eq48}
\mathbf{P_c}(n)&=&1 - \frac{1}{k} \cdot \left[2(k-1)\cdot\left(\frac{1}{4}-\frac{1}{8\lfloor n/2 \rfloor +4}\right)\right]\nonumber\\
&=&\frac{k+1}{2k} + \frac{k-1}{k} \frac{1}{4\lfloor n/2 \rfloor +2}.
\end{eqnarray}
\end{proof}

(3) Detection probability when $\delta > 3$

When $n$ is finite, we can numerically compute the exact detection probability using Algorithm 2 in Section IV. When $n \rightarrow \infty$, we can obtain the asymptotic correct detection probability using Lemma 7 and an insightful analysis into the graph structure.

\begin{proof}[Proof of Theorem 3-{\romannumeral3}]
{From \eqref{eq341}, we have
\begin{equation}
\label{eq52}
\mathbf{P_G}(E_1)=\mathbf{P_G}(F_1)
=I_{1/2}\left(\frac{1}{\delta-2},\frac{\delta-1}{\delta-2}\right)+\xi_(n,\delta),
\end{equation}
where $\lim_{n \rightarrow \infty}\xi_(n,\delta)=0$.
}

Using Lemma 7, the correct detection probability for a suspect node $s^\ast$ with $m$ neighbors in the suspect set $S$ is
{\begin{equation}
\label{eq53}
\mathbf{P_c}(n|s^\ast)=1-m\left(1-I_{1/2}\left(\frac{1}{\delta-2},\frac{\delta-1}{\delta-2}\right)-\xi(n,\delta)\right).
\end{equation}
Note that the residual term $\xi(n,\delta)$ is identical\footnotemark[2] for all $m$ neighboring suspect nodes of $s^\ast$.
}
\footnotetext [2] {Due to the following Lemma 9, all events $E^c_j=\{X_j \geq {n}/{2}\}, 1 \leq j \leq m$ are disjoint and symmetric, hence so are $F^c_j=\{X_j > {n}/{2}\}, 1 \leq j \leq m$.}

For a suspect set $S$ with cardinality $k$ that forms a connected subgraph of a regular tree $G$ with node degree $\delta>3$, for each suspect node $s^\ast \in S$, we first find the number of its neighboring suspect nodes. Note that, {$\mathbf{P_c}(n|s^\ast)$ is one subtracted by a common factor $m$ times,} each of which accounting for one neighboring suspect node of $s^\ast$ connected by an edge. Since there are $k-1$ edges connecting the $k$ suspect nodes in $S$, each edge will account for a reduction of the factor {\it twice}. Therefore, from \eqref{eq39}, we have

\small
\begin{eqnarray}
\label{eq54}
\mathbf{P_c}(n)
&=&1-\frac{1}{k} \cdot \left[{2(k-1)\cdot\left(1-I_{1/2}\left(\frac{1}{\delta-2},\frac{\delta-1}{\delta-2}\right)-\xi(n)\right)}\right]\nonumber\\
&=&1-\frac{2(k-1)}{k}+\frac{2(k-1)}{k}I_{1/2}\left(\frac{1}{\delta-2},\frac{\delta-1}{\delta-2}\right)\nonumber\\
&&+\frac{2(k-1)}{k}\xi(n,\delta).
\end{eqnarray}

\normalsize
From Fig. 6, we see that $I_{1/2}\left(\frac{1}{\delta-2},\frac{\delta-1}{\delta-2}\right) > 0.75$ as the node degree $\delta > 3$. Therefore, we have
\begin{equation}
\label{eq55}
\mathbf{P_c}(n) > \frac{k+1}{2k},
\end{equation}
{for any sufficiently large $n$ and $\delta > 3$.}

Besides, since $2(k-1)/k=2$ as $k \rightarrow \infty$, thus from \eqref{eq54}, we have
\begin{equation}
\label{eq560}
\lim_{n \rightarrow \infty} \mathbf{P_c}(n) = 2I_{1/2}\left(\frac{1}{\delta-2},\frac{\delta-1}{\delta-2}\right)-1.
\end{equation}

Furthermore, since $I_{1/2}\left(0,1\right) = 1$ (setting $\delta \rightarrow \infty$), we thus have
\begin{equation}
\label{eq56}
\lim_{\delta \rightarrow \infty} \lim_{n \rightarrow \infty}\mathbf{P_c}(n)\rightarrow 1.
\end{equation}
Importantly, note that the growth of $\delta$ does not need to depend on the growth of $n$.

\end{proof}

\subsection{Proof of Theorem 4: Two Suspects}

In the case where $S=\{s_1,s_2\}$ contains only two suspect nodes, let $d$ be the shortest path distance between $s_1$ and $s_2$ on $G$. We assume $s_1$ to be the rumor source $s^\ast$ by symmetry, let $\mathcal{P}=\{v_0=s_1,v_1,\cdots,v_d=s_2\}$ be the shortest path from $s_1$ to $s_2$, and define a random variable $Z_h$ to be the number of nodes in the subtree $T_{v_h}^{s^\ast}$ ($1 \leq h \leq d$). It is clear that $Z_h \geq Z_{h+1}+1$ if $Z_h>0$ for all $1 \leq h \leq d-1$.

In the following, we focus on the error detection probability, i.e., $\mathbf{P_e}(n)=1-\mathbf{P_c}(n)$. As a result, we assume $Z_h>0$ for all $1 \leq h \leq d$. Since there are only two suspect nodes in $S$, $\mathbf{P_e}(n)$ can be written as
\begin{eqnarray}
\label{eq631}
\mathbf{P_e}(n)&=&\frac{1}{2} \cdot \sum_{R(s^\ast,G_n) = R(s_2,G_n)}{\mathbf{P_G}}\left[\bigcap_{h=1}^{d}(Z_h=z_h)\right]\nonumber\\
&&+\sum_{R(s^\ast,G_n) < R(s_2,G_n)}{\mathbf{P_G}}\left[\bigcap_{h=1}^{d}(Z_h=z_h)\right],
\end{eqnarray}
where ${\mathbf{P_G}}\left[\bigcap_{h=1}^{d}(Z_h=z_h)\right]$ is given by \eqref{eq62}.

When $n$ is finite, we can numerically compute the exact detection probability using Algorithm 3 in Section IV. In the following, we will establish the argument of Theorem 4.

(1) Detection probability when $\delta=2$

\begin{proof}[Proof of Theorem 4-{\romannumeral1}]
From \eqref{eq6}, we have $R(s^\ast,G_n)=\binom{n-1}{z_1}$ and $R(s_2,G_n)=\binom{n-1}{z_1-d}$. If $R(s^\ast,G_n) \leq R(s_2,G_n)$, then $z_1 \geq (n+d+1)/2$. Note that we only need to consider the distribution of $Z_1$. When $\delta=2$, the distribution in \eqref{eq60} can be written as
\begin{equation}
\label{eq63}
{\mathbf{P_G}}\left[Z_1=z_1\right]=\frac{(n-1)!}{z_1! (n-z_1-1)!} \frac{1}{2^{n-1}}.
\end{equation}

From \eqref{eq631}, the error detection probability is
\begin{eqnarray}
\label{eq64}
\mathbf{P_e}(n)&=&\frac{1}{2}\sum_{z_1=(n+d+1)/2}{\mathbf{P_G}}\left[Z_1=z_1\right]
+\sum_{z_1>(n+d+1)/2}{\mathbf{P_G}}\left[Z_1=z_1\right],\nonumber\\
&=&\begin{cases}
\begin{aligned}
\frac{1}{2}-\frac{1}{2^n}\sum_{z_1=(n-d-1)/2}^{(n+d+1)/2}{\binom{n-1}{z_1}}, \, (n-d) \, \mbox{is odd};
\end{aligned}\\
\begin{aligned}
\frac{1}{2}-\frac{1}{2^n}\sum_{z_1=(n-d)/2}^{(n+d-2)/2}{\binom{n-1}{z_1}}, \, (n-d) \, \mbox{is even}.
\end{aligned}
\end{cases}
\end{eqnarray}
In the above, the detailed deduction is in Appendix-G.
\end{proof}

(2) Detection probability when $\delta=3$

\begin{proof}[Proof of Theorem 4-{\romannumeral2}]
First of all, consider the case when $d=1$. From \eqref{eq47}, the error detection probability is
\begin{eqnarray}
\label{eq65}
\mathbf{P_e}(n)&=&1-\mathbf{P_c}(n)\nonumber\\
&=&\frac{1}{4}-\frac{1}{4} \frac{1}{2\lfloor n/2 \rfloor +1}.
\end{eqnarray}

Now, consider the case when $d=2$. The distribution in \eqref{eq62} can be written as
\begin{equation}
\label{eq66}
{\mathbf{P_G}}\left[\bigcap_{h=1}^{2}(Z_h=z_h)\right]=\frac{2(n-z_1)}{n(n+1)z_1}.
\end{equation}

From \eqref{eq7}, we have $R(s_2,G_n)=R(s^\ast,G_n)\frac{z_1}{n-z_1}\frac{z_2}{n-z_2}$. If $R(s^\ast,G_n) \leq R(s_2,G_n)$, then $z_1+z_2 \geq n$. Using \eqref{eq631} and letting $n \rightarrow \infty$, the error detection probability is
\small
\begin{eqnarray}
\label{eq67}
\mathbf{P_e}(n)&=&\frac{1}{2}\sum_{z_1+z_2=n}{\mathbf{P_G}}\left[\bigcap_{h=1}^{2}(Z_h=z_h)\right]
+\sum_{z_1+z_2>n}{\mathbf{P_G}}\left[\bigcap_{h=1}^{2}(Z_h=z_h)\right]\nonumber\\
&\approx&0.114.
\end{eqnarray}
\normalsize
In the above, the detailed deduction is in Appendix-H.
\end{proof}

(3) Detection probability when $\delta > 3$

\begin{proof}[Proof of Theorem 4-{\romannumeral3}]
Consider the case when $d=1$. From \eqref{eq53}, the error detection probability is
\begin{eqnarray}
\label{eq70}
\mathbf{P_e}(n)&=&1-\mathbf{P_c}(n)\nonumber\\
&=&1-I_{1/2}\left(\frac{1}{\delta-2},\frac{\delta-1}{\delta-2}\right).
\end{eqnarray}
as $n \rightarrow \infty$.

Furthermore, since $I_{1/2}\left(0,1\right) = 1$ (setting $\delta \rightarrow \infty$), we thus have
\begin{equation}
\label{eq71}
\lim_{\delta \rightarrow \infty} \lim_{n \rightarrow \infty} \mathbf{P_e}(n)=0.
\end{equation}
\end{proof}

(4) Increasing detection probability with distance

\begin{proof}[Proof of Theorem 4-{\romannumeral4}]
Equivalently, we prove that $\mathbf{P_e}(n)$ decreases with $d$. Without loss of generality, we assume $Z_h>0$ for all $1 \leq h \leq d$. The proof will be completed by showing that if $R(v_d,G_n) \geq R(s^\ast,G_n)$ then $R(v_{d-1},G_n) > R(s^\ast,G_n)$ for all $d \geq 2$, which is to be verified by contradiction to the assumption $R(v_d,G_n) \geq R(s^\ast,G_n)$.

Suppose $R(v_{d-1},G_n) \leq R(s^\ast,G_n)$, then $Z_{d-1} \leq n/2$. Otherwise, $Z_{d-1} > n/2$ and thus $Z_h > n/2$ for all $1 \leq h \leq d-1$; namely, ${Z_h}/{(n-Z_h)} > 1$ for all $1 \leq h \leq d-1$. Repeatedly using \eqref{eq7}, we have
\begin{equation}
R(v_{d-1},G_n)=R(s^\ast,G_n)\frac{Z_1}{n-Z_1}\frac{Z_2}{n-Z_2}\cdots\frac{Z_{d-1}}{n-Z_{d-1}},
\end{equation}
which leads to the contradiction that $R(v_{d-1},G_n)>R(s^\ast,G_n)$. As a result, we have $Z_{d-1} \leq n/2$.

As $Z_{d-1} \leq n/2$, then $Z_d < n/2$ and thus ${Z_d}/{(n-Z_d)} < 1$. As a result, we have
\begin{equation}
R(v_d,G_n) = R(v_{d-1},G_n)\frac{Z_d}{n-Z_d} < R(v_{d-1},G_n) \leq R(s^\ast,G_n),
\end{equation}
which is in contradiction to the assumption of $R(v_d,G_n) \geq R(s^\ast,G_n)$.
\end{proof}

\subsection{Proof of Theorem 5: Multiple Suspects}

Consider the case where $S=\{s_1,s_2,\ldots,s_k\}$ with cardinality $k$ forms a general subgraph of the network $G$. From \eqref{eq39}, we have
\begin{equation}
\label{eq77}
\mathbf{P_c}(n)=\sum_{i=1}^{k}\mathbf{P_s}(s_i)\mathbf{P_c}(n|s_i)=\frac{1}{k}\sum_{s^\ast \in S}\mathbf{P_c}(n|s^\ast).
\end{equation}

We first find the correct detection probability $\mathbf{P_c}(n|s^\ast)$ for each suspect node $s^\ast \in S$. Assume that $s^\ast \in S$ is the rumor source and it has $m$ ($m \leq \delta$) neighbors $N_l(s^\ast)=\{v_1,\ldots,v_m\}$, subject to that each subtree $T_{v_j}^{s^\ast}$ ($1 \leq j \leq m$) of $G$ contains at least one suspect node in $S$. In the following, we call the nodes in $N_l(s^\ast)$ as {\it suspect neighbors}. Let a random variable $X_j$ be the number of nodes in each subtree $T_{v_j}^{s^\ast}$ ($1 \leq j \leq m$) of $G_n$, then we have the following lemma; for its proof see Appendix-I.

\begin{lemma}
To correctly identify source $s^\ast$ with $m$ suspect neighbors, we have
\begin{equation}
\label{eq78}
\mathbf{P_c}(n|s^\ast) \geq 1-\mathbf{P_c}(n|s^\ast, \mbox{conn}),
\end{equation}
where $\mathbf{P_c}(n|s^\ast, \mbox{conn})$ is the correct detection probability for $s^\ast$ when all the $m$ suspect neighbors are neighboring suspect nodes of $s^\ast$.
\end{lemma}

\emph{Remark 8:} Lemma 8 is deduced from Proposition 1. In order to prove Theorem 5, we shall find the lower bound of $\mathbf{P_c}(n|s^\ast)$ in the following.

When the $k$ suspect nodes in $S$ form a connected subgraph of the network $G$, we can exactly compute $\mathbf{P_c}(n)$ in both the finite and asymptotic regimes. In the following, we will establish the argument of Theorem 5.

(1) Detection probability when $\delta=2$
\begin{proof}[Proof of Theorem 5-{\romannumeral1}]
We use Lemma 8 and \eqref{eq43} to obtain a lower bound of the correct detection probability for a suspect node $s^\ast$ with $m$ suspect neighbors:
\begin{equation}
\label{eq79}
\mathbf{P_c}(n|s^\ast)\geq
\begin{cases}
\begin{aligned}
\frac{1}{2} + \frac{1}{2^n} \binom{n-1}{\lfloor(n-1)/2\rfloor},m=1;
\end{aligned}\\
\begin{aligned}
\frac{1}{2^{n-1}} \binom{n-1}{\lfloor(n-1)/2\rfloor},m=2.
\end{aligned}
\end{cases}
\end{equation}

For a suspect set $S$ with cardinality $k$ that forms a general subgraph of a linear network $G$, we know that only the two suspect nodes at the endpoints of the sub-linear network have one suspect neighbor and all other suspect nodes have two suspect neighbors. Therefore, from \eqref{eq77}, we have
\begin{equation}
\label{eq80}
\mathbf{P_c}(n) \geq \frac{1}{k}\left[1 + \frac{k-1}{2^{n-1}} \binom{n-1}{\lfloor(n-1)/2\rfloor}\right].
\end{equation}

As $n \rightarrow \infty$, by the Stirling's formula, we have
\begin{equation}
\label{eq81}
\mathbf{P_c}(n) \geq \frac{1}{k}+\frac{k-1}{k}\mathcal{O}\left(\frac{1}{\sqrt{n}}\right)=\frac{1}{k}+\mathcal{O}\left(\frac{1}{\sqrt{n}}\right).
\end{equation}
\end{proof}

(2) Detection probability when $\delta=3$

\begin{proof}[Proof of Theorem 5-{\romannumeral2}]
We use Lemma 8 and \eqref{eq47} to obtain a lower bound of the correct detection probability for a suspect node $s^\ast$ with $m$ suspect neighbors as
\begin{equation}
\label{eq82}
\mathbf{P_c}(n|s^\ast)\geq
\begin{cases}
\begin{aligned}
\frac{3}{4}+\frac{1}{4} \frac{1}{2\lfloor n/2 \rfloor +1},m=1;
\end{aligned}\\
\begin{aligned}
\frac{1}{2}+\frac{1}{2} \frac{1}{2\lfloor n/2 \rfloor +1},m=2;
\end{aligned}\\
\begin{aligned}
\frac{1}{4}+\frac{3}{4} \frac{1}{2\lfloor n/2 \rfloor +1},m=3.
\end{aligned}
\end{cases}
\end{equation}

For a suspect set $S$ with cardinality $k$ that forms a general subgraph of a regular tree $G$ with node degree $\delta=3$, for each suspect node $s^\ast \in S$, we first find the number of its suspect neighbors. Note that given $s^\ast$ with $m$ suspect neighbors, the lower bound of $\mathbf{P_c}(n|s^\ast)$ is one subtracted by a same factor, $1/4-1/(8\lfloor n/2 \rfloor +4)$, $m$ times, each of which accounting for one suspect neighbors of $s^\ast$ connected by an edge. Since there are at most $2(k-1)$ edges connecting the $k$ suspect nodes in $S$ and their suspect neighbors, each edge will account for a reduction of the factor once. Therefore, from \eqref{eq77}, we have
\begin{equation}
\label{eq83}
\mathbf{P_c}(n) \geq \frac{k+1}{2k} + \frac{k-1}{k} \frac{1}{4\lfloor n/2 \rfloor +2}.
\end{equation}
\end{proof}

(3) Detection probability when $\delta > 3$
\begin{proof}[Proof of Theorem 5-{\romannumeral3}]
We use Lemma 8 and \eqref{eq53} to obtain a lower bound of the correct detection probability for a suspect node $s^\ast$ with $m$ suspect neighbors as
{
\begin{equation}
\label{eq84}
\mathbf{P_c}(n|s^\ast) \geq 1-m\left(1-I_{1/2}\left(\frac{1}{\delta-2},\frac{\delta-1}{\delta-2}\right)-\xi(n,\delta)\right).
\end{equation}
}

For a suspect set $S$ with cardinality $k$ that forms a general subgraph of a regular tree $G$ with node degree $\delta>3$, for each suspect node $s^\ast \in S$, we first find the number of its suspect neighbors. {Note that the lower bound of $\mathbf{P_c}(n|s^\ast)$ is one subtracted by a common factor $m$ times,} each of which accounting for one suspect neighbors of $s^\ast$ connected by an edge. Since there are at most $2(k-1)$ edges connecting the $k$ suspect nodes in $S$ and their suspect neighbors, each edge will account for a reduction of the factor {once. Therefore,} from \eqref{eq77}, we have

{
\small
\begin{eqnarray}
\label{eq85}
\mathbf{P_c}(n) \geq 1-\frac{2(k-1)}{k}+\frac{2(k-1)}{k}I_{1/2}\left(\frac{1}{\delta-2},\frac{\delta-1}{\delta-2}\right)+\frac{2(k-1)}{k}\xi(n,\delta).
\end{eqnarray}
}

From Fig. 6, we see that $I_{1/2}\left(\frac{1}{\delta-2},\frac{\delta-1}{\delta-2}\right) > 0.75$ for $\delta > 3$. Therefore, we have
{\begin{equation}
\label{eq86}
\mathbf{P_c}(n) > \frac{k+1}{2k},
\end{equation}
for any sufficiently large $n$ and $\delta > 3$.
}

Besides, since $2(k-1)/k=2$ as $k \rightarrow \infty$, thus from \eqref{eq85}, we have
\begin{equation}
\label{eq870}
\lim_{n \rightarrow \infty} \mathbf{P_c}(n) \geq 2I_{1/2}\left(\frac{1}{\delta-2},\frac{\delta-1}{\delta-2}\right)-1.
\end{equation}

Furthermore, since $I_{1/2}\left(0,1\right) = 1$ (setting $\delta \rightarrow \infty$) we thus have
\begin{equation}
\label{eq87}
\lim_{\delta \rightarrow \infty} \lim_{n \rightarrow \infty} \mathbf{P_c}(n)=1.
\end{equation}
Importantly, note that the growth of $\delta$ does not need to depend on the growth of $n$.

\end{proof}

(4) The {worst-case detection} probability

Before the argument of Theorem 5-{\romannumeral4}, we present the following lemma, which will be also used in the design of the numerical algorithms in Section IV; for its proof see Appendix-J. For a source $s^\ast$ with $m$ ($m \leq \delta$) neighbors $N_l(s^\ast)=\{v_1,\ldots,v_m\}$ in $S$, let a random variable $X_j$ be the number of nodes in each subtree $T_{v_j}^{s^\ast}$ ($1 \leq j \leq m$) of $G_n$.

\begin{lemma}
To correctly identify source $s^\ast$ with $m$ neighboring suspect nodes as the estimate $\hat{s}$, we have
\begin{equation}
\label{eq871}
\mathbf{P_c}(n|s^\ast)=1-m\left(0.5{\mathbf{P_G}}\left(X_1=\frac{n}{2}\right)+\sum_{x_1>n/2}{\mathbf{P_G}}\left(X_1=x_1\right)\right),
\end{equation}
where ${\mathbf{P_G}}\left(X_1=x_1\right)$ is given by \eqref{eq240}.
\end{lemma}

\emph{Remark 9:} Lemma 9 is deduced from Proposition 1. In fact, we can use Lemma 9 to deduce the exact detection probability $\mathbf{P_c}(n)$ in the finite regime for Theorems 2 and 3, instead of using Lemma 6.

\begin{proof}[Proof of Theorem 5-{\romannumeral4}]
First, consider the case when the $k$ suspect nodes are connected, and we use Lemma 9 to derive the minimum of $\mathbf{P_c}(n)$.

For a suspect set $S$ with cardinality $k$ that forms a connected subgraph of a regular tree $G$ with node degree $\delta \geq 2$, for each suspect node $s^\ast \in S$, we first find the number of its neighboring suspect nodes. Note that $\mathbf{P_c}(n|s^\ast)$ in \eqref{eq871} is one subtracted by a common factor $m$ times, each of which accounting for one neighboring suspect node of $s^\ast$ connected by an edge. Since there are $k-1$ edges connecting the $k$ suspect nodes in $S$, each edge will account for a reduction of the factor twice. Therefore, from \eqref{eq77}, we have
\begin{equation}
\label{eq872}
\mathbf{P_c}(n)=1-\frac{2(k-1)}{k}\left(0.5{\mathbf{P_G}}\left(X_1=\frac{n}{2}\right)+\sum_{x_1>n/2}{\mathbf{P_G}}\left(X_1=x_1\right)\right).
\end{equation}

Next, consider the case when the $k$ suspect nodes form a general subgraph of $G$. From Lemmas 8 and 9, a lower bound of the correct detection probability for a suspect node $s^\ast$ with $m$ suspect neighbors is
\begin{equation}
\label{eq873}
\mathbf{P_c}(n|s^\ast) \geq 1-m\left(0.5{\mathbf{P_G}}\left(X_1=\frac{n}{2}\right)+\sum_{x_1>n/2}{\mathbf{P_G}}\left(X_1=x_1\right)\right).
\end{equation}

For a suspect set $S$ with cardinality $k$ that forms a general subgraph of a regular tree $G$ with node degree $\delta \geq 2$, for each suspect node $s^\ast \in S$, we first find the number of its suspect neighbors. {Note that the lower bound of $\mathbf{P_c}(n|s^\ast)$ is one subtracted by a common factor $m$ times,} each of which accounting for one suspect neighbors of $s^\ast$ connected by an edge. Since there are at most $2(k-1)$ edges connecting the $k$ suspect nodes in $S$ and their suspect neighbors, each edge will account for a reduction of the factor once. Therefore, from \eqref{eq77}, we have

\small
\begin{eqnarray}
\label{eq874}
\mathbf{P_c}(n) \geq 1-\frac{2(k-1)}{k}\left(0.5{\mathbf{P_G}}\left(X_1=\frac{n}{2}\right)+\sum_{x_1>n/2}{\mathbf{P_G}}\left(X_1=x_1\right)\right),
\end{eqnarray}

\normalsize
\noindent where this lower bound is achievable if and only if the $k$ suspect nodes are connected.
\end{proof}

\section{Detection Probability on Regular Trees: Numerical Computation}

In this section, we propose algorithms to compute the exact detection probabilities of the MAP estimator in the finite regime for regular tree-type networks, based on the theoretical results established in Section III. We focus on three representative scenarios: First, the suspect set contains all the nodes in the network; Second, the suspect set forms {a connected subgraph} of the network; Third, the suspect set contains only two nodes. Their corresponding algorithms are listed in Table II. We further establish that the correct detection probability monotonically decreases with the number of infected nodes and monotonically increases with the node degree for the former two cases.

\begin{table}[ht]
  \centering
  \caption{Algorithms for Computing the Detection Probability of the MAP Estimator}
  \begin{tabular}{|l|c|} \hline
  Suspect characteristics & Numerical computation\\ \hline
  All nodes are suspects & Algorithm 1\\ \hline
  Suspects forms a connected subgraph & Algorithm 2\\ \hline
  Two suspects & Algorithm 3\\
  \hline\end{tabular}
\end{table}

\subsection{Suspecting all Nodes}

\begin{algorithm}[b]
\caption{AllSuspect($\delta,n$)}
\label{alg1}
\begin{algorithmic}
\State $\mbox{Initialize} \, \mathbf{P_c}(n)=0$
\If {$\delta=2$}
    \State $\mbox{Compute} \, \mathbf{P_c}(n) \, \mbox{using \eqref{eq21}}$
\ElsIf {$\delta=3$}
    \State $\mbox{Compute} \, \mathbf{P_c}(n) \, \mbox{using \eqref{eq22}}$
\Else
    \State $tmp=0.5\cdot{\mathbf{P_G}}\left(X_1=n/2\right)$
    \For {$x_1=\lfloor{n/2}\rfloor+1 \to n-1$}
        \State $tmp=tmp + {\mathbf{P_G}}\left(X_1=x_1\right)$
    \EndFor
    \State $\mathbf{P_c}(n)=1-\delta \cdot tmp$
\EndIf
\State $\mbox{Output} \, \mathbf{P_c}(n)$
\end{algorithmic}
\end{algorithm}


In this case, $S=V$, we only need to consider an arbitrary node $s^\ast \in G$ as the rumor source by symmetry. For a source $s^\ast$ with $\delta$ neighbors $N_l(s^\ast)=\{v_1,\ldots,v_\delta\} \subset S$, let a random variable $X_j$ be the number of nodes in each subtree $T_{v_j}^{s^\ast}$ ($1 \leq j \leq \delta$) of $G_n$. Then, we use Algorithm 1 to compute the exact detection probability for this case.

When $\delta=2$ and $\delta=3$, we use \eqref{eq21} and \eqref{eq22} to compute $\mathbf{P_c}(n)$, respectively. When $\delta>3$, from Lemma 9, we have
\begin{equation}
\label{eq88}
\mathbf{P_c}(n)=1-\delta\left(0.5{\mathbf{P_G}}\left(X_1=\frac{n}{2}\right)+\sum_{x_1>n/2}{\mathbf{P_G}}\left(X_1=x_1\right)\right).
\end{equation}

Based on \eqref{eq88}, we have the following corollary; for its proof see Appendix-K.

\begin{corollary}
Suppose $S=V$, i.e., every infected node is a suspect node, then:\\
{\romannumeral1}) $\mathbf{P_c}(n)$ monotonically decreases with $n$;\\
{\romannumeral2}) $\mathbf{P_c}(n)$ monotonically increases with $\delta$.
\end{corollary}

\emph{Remark 10:} Corollary 10 is complementary to Theorem 2. The MAP estimator has better detection performance when {$n$ is smaller}, i.e., at an earlier stage of the rumor spreading process. On the other hand, the MAP estimator performs better for a network with a richer connectivity (higher node degree).

\subsection{Connected Suspects}

\begin{algorithm}[t]
\caption{ConnSuspect($\delta,k,n$)}
\label{alg2}
\begin{algorithmic}
\State $\mbox{Initialize} \, \mathbf{P_c}(n)=0$
\If {$\delta=2$}
    \State $\mbox{Compute} \, \mathbf{P_c}(n) \, \mbox{using \eqref{eq36}}$
\ElsIf {$\delta=3$}
    \State $\mbox{Compute} \, \mathbf{P_c}(n) \, \mbox{using \eqref{eq37}}$
\Else
    \State $tmp=0.5\cdot{\mathbf{P_G}}\left(X_1=n/2\right)$
    \For {$x_1=\lfloor{n/2}\rfloor+1 \to n-1$}
        \State $tmp=tmp + {\mathbf{P_G}}\left(X_1=x_1\right)$
    \EndFor
    \State $\mathbf{P_c}(n)=1-\frac{2(k-1)}{k} \cdot tmp$
\EndIf
\State $\mbox{Output} \, \mathbf{P_c}(n)$
\end{algorithmic}
\end{algorithm}


In this case, $S=\{s_1,s_2,\ldots,s_k\}$ with cardinality $k$ forms a connected subgraph of the network $G$. Assume that $s^\ast \in S$ is the rumor source and it has $m$ ($m \leq \delta$) neighbors $N_l(s^\ast)=\{v_1,\ldots,v_m\} \subset S$. Let a random variable $X_j$ be the number of nodes in each subtree $T_{v_j}^{s^\ast}$ ($1 \leq j \leq m$) of $G_n$. Then, we use Algorithm 2 to compute the exact detection probability for this case.

When $\delta=2$ and $\delta=3$, we use \eqref{eq36} and \eqref{eq37} to compute $\mathbf{P_c}(n)$, respectively. When $\delta>3$, from Lemma 9 and \eqref{eq872}, we have
\begin{equation}
\label{eq91}
\mathbf{P_c}(n)=1-\frac{2(k-1)}{k}\left(0.5{\mathbf{P_G}}\left(X_1=\frac{n}{2}\right)+\sum_{x_1>n/2}{\mathbf{P_G}}\left(X_1=x_1\right)\right).
\end{equation}

Based on \eqref{eq91}, we have the following corollary which is complementary to Theorem 3; its proof is similar to that of Corollary 10 and hence omitted.

\begin{corollary}
Suppose that $S$ forms a connected subgraph of $G$, then:\\
{\romannumeral1}) $\mathbf{P_c}(n)$ monotonically decreases with $n$;\\
{\romannumeral2}) $\mathbf{P_c}(n)$ monotonically increases with $\delta$.
\end{corollary}

\subsection{Two Suspects}

\begin{algorithm}[t]
\caption{TwoSuspect($\delta,d,n$)}
\label{alg3}
\begin{algorithmic}
\State $\mbox{Initialize} \, \mathbf{P_c}(n)=0$
\If {$\delta=2$}
    \State $\mbox{Compute} \, \mathbf{P_c}(n) \, \mbox{by \eqref{eq57}}$
\Else
    \State $tmp=0$
    \State $\mbox{Enumerate} \, \Phi=\{G_n \mid R(s^\ast,G_n)=R(s_2,G_n)\}$
    \For {$\mbox{each infection} \, G_n \in \Phi$}
        \State $tmp=tmp + 0.5\cdot{\mathbf{P_G}}\left[\bigcap_{h=1}^{d}(Z_h=z_h)\right]$
    \EndFor
    \State $\mbox{Enumerate} \, \Phi=\{G_n \mid R(s^\ast,G_n)<R(s_2,G_n)\}$
    \For {$\mbox{each infection} \, G_n \in \Phi$}
        \State $tmp=tmp + {\mathbf{P_G}}\left[\bigcap_{h=1}^{d}(Z_h=z_h)\right]$
    \EndFor
    \State $\mathbf{P_c}(n)=1-tmp$
\EndIf
\State $\mbox{Output} \, \mathbf{P_c}(n)$
\end{algorithmic}
\end{algorithm}


In this case, $S=\{s_1,s_2\}$ contains only two suspect nodes. Let $d$ be the shortest path distance between $s_1$ and $s_2$ in $G$. We assume $s_1$ to be the rumor source $s^\ast$ by symmetry, let $\mathcal{P}=\{v_0=s_1,v_1,\cdots,v_d=s_2\}$ be the shortest path from $s_1$ to $s_2$, and define a random variable $Z_h$ to be the number of nodes in the subtree $T_{v_h}^{s^\ast}$ ($1 \leq h \leq d$). Then, we have Algorithm 3 to compute the exact detection probability for this case.

When $\delta=2$, we use \eqref{eq57} to compute $\mathbf{P_c}(n)$. When $\delta \geq 3$, from \eqref{eq631}, we have
\begin{eqnarray}
\label{92}
\mathbf{P_c}(n)&=&1-\frac{1}{2} \cdot \sum_{R(s^\ast,G_n) = R(s_2,G_n)}{\mathbf{P_G}}\left[\bigcap_{h=1}^{d}(Z_h=z_h)\right]\nonumber\\
&&-\sum_{R(s^\ast,G_n) < R(s_2,G_n)}{\mathbf{P_G}}\left[\bigcap_{h=1}^{d}(Z_h=z_h)\right],
\end{eqnarray}
where ${\mathbf{P_G}}\left[\bigcap_{h=1}^{d}(Z_h=z_h)\right]$ is given by \eqref{eq62}.

\section{Numerical Experiments}

\begin{figure}[t]
\label{fig7}
\center
  \includegraphics[width=0.5\textwidth]{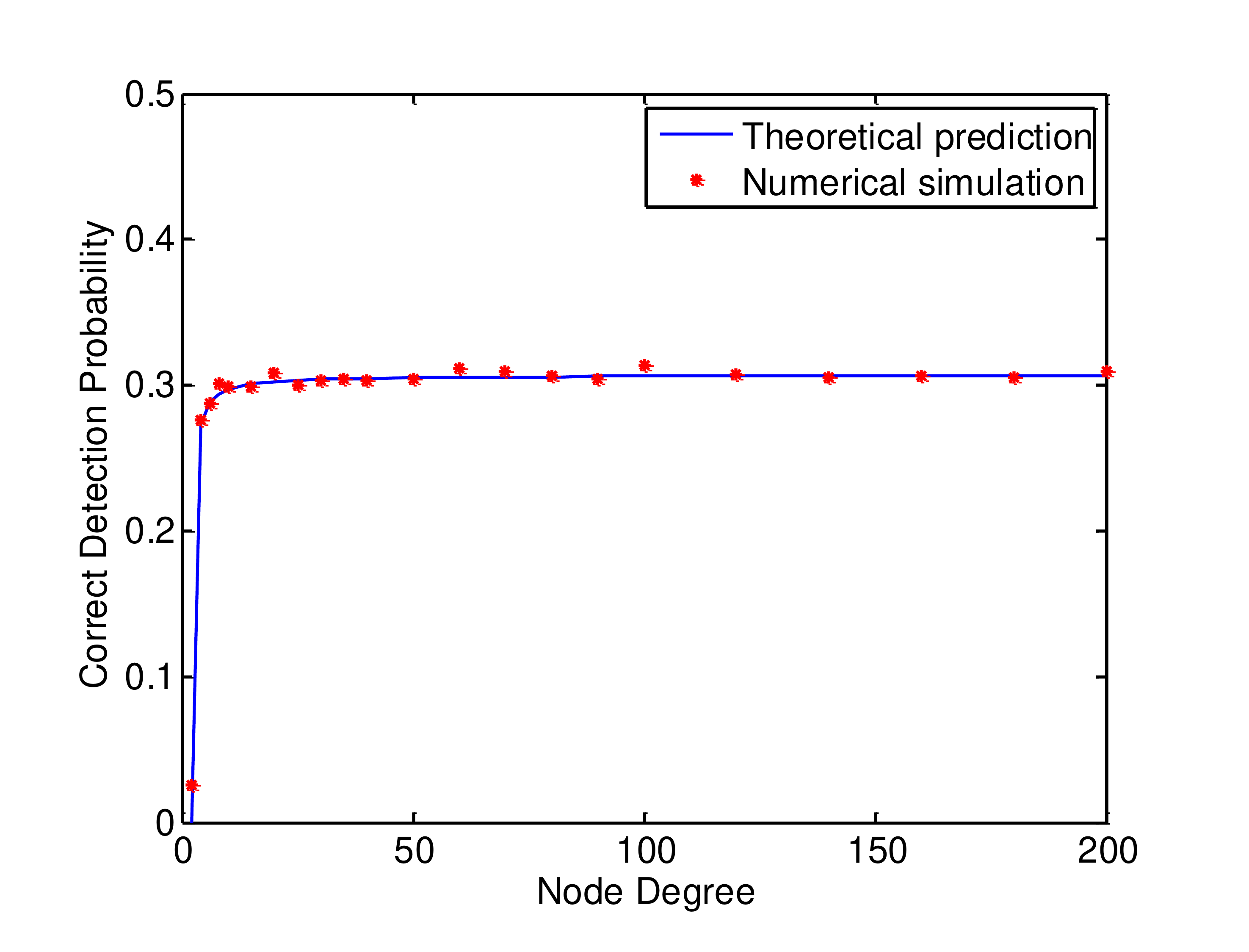}\\
  \caption{Detection probability when $S=V$.}
\end{figure}

\begin{figure}[ht]
\label{fig8}
\center
  \includegraphics[width=0.5\textwidth]{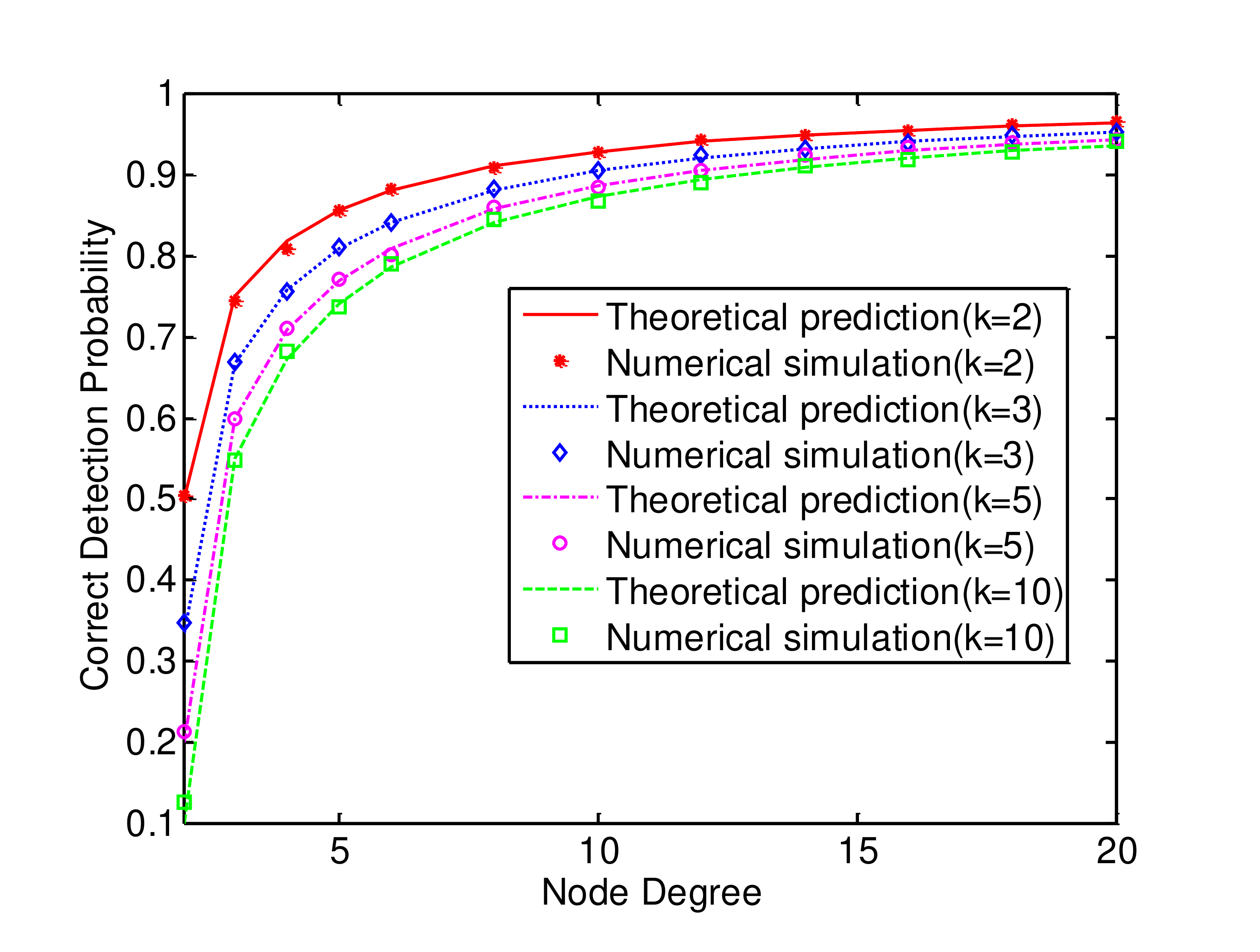}\\
  \caption{Detection probability when $S$ forms a connected subgraph of $G$.}
\end{figure}

\begin{figure}[t]
\label{fig9}
\center
  \includegraphics[width=0.5\textwidth]{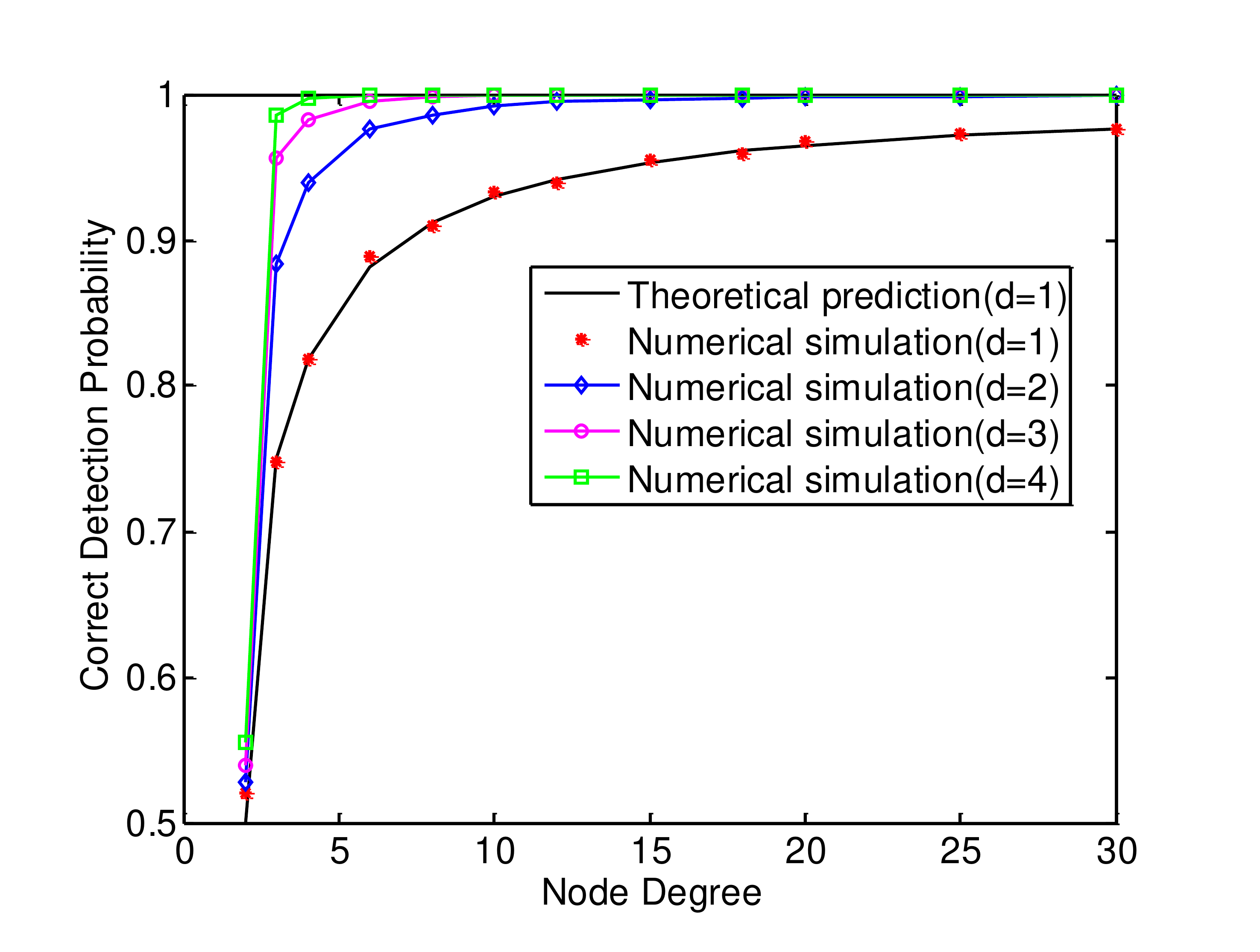}\\
  \caption{Detection probability when $S$ contains two suspect nodes.}
\end{figure}

\begin{figure}[t]
\label{fig10}
\center
  \includegraphics[width=0.5\textwidth]{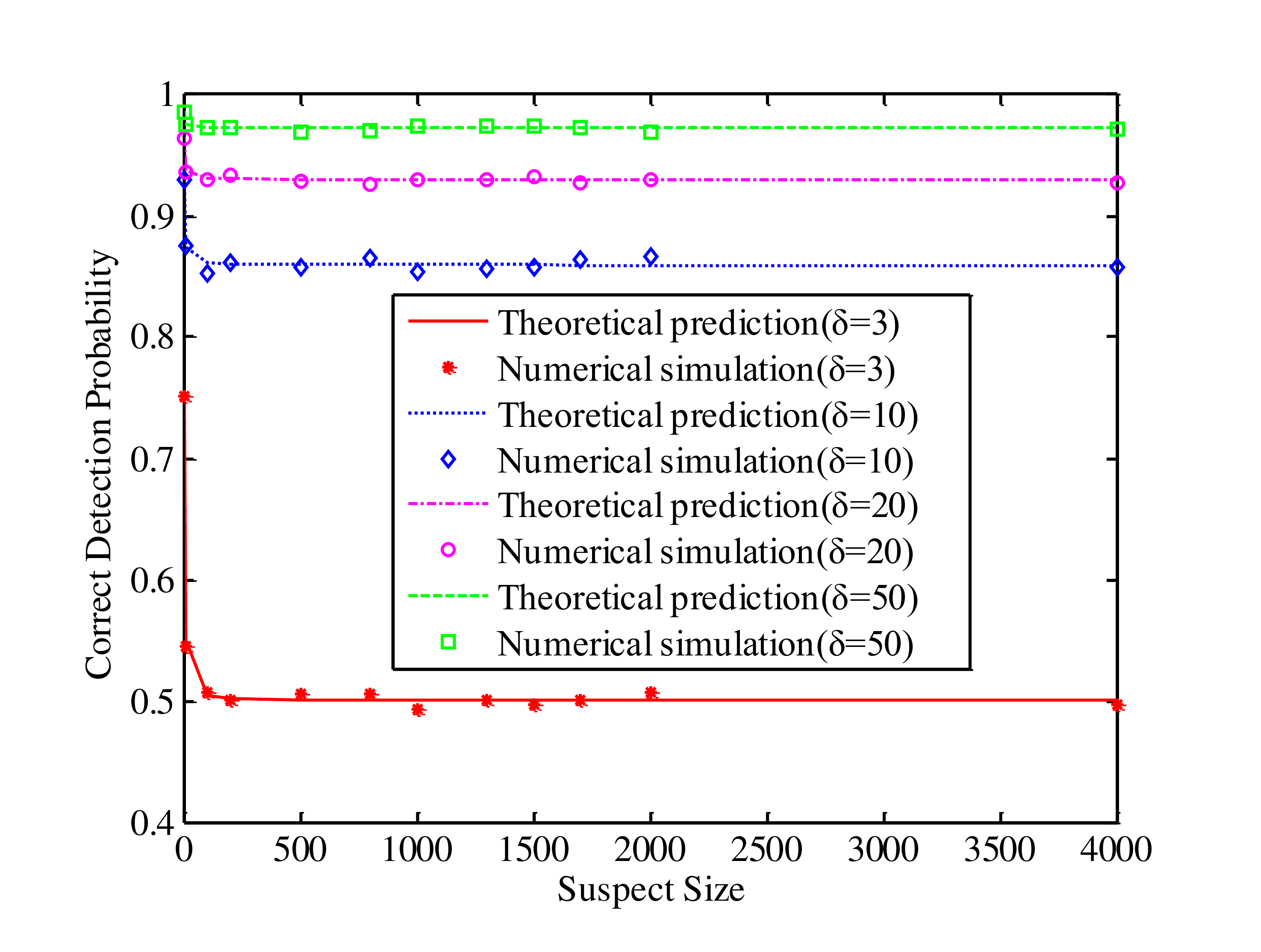}\\
  \caption{{Detection probability when $S$ forms a connected subgraph of $G$.}}
\end{figure}

In this section, we carry out simulation experiments to corroborate and illustrate our analysis. For verifying the asymptotic results, in each experiment run we let $n = 1000$ nodes be eventually infected by a rumor source node uniformly randomly chosen from the suspect set, following the SI model, and use the estimator \eqref{eq3} to identify this source.

For the first scenario of $S = V$, it is shown in Fig. 7 that the correct detection probability is increasing with the node degree $\delta$, from virtually zero when $\delta=2$ to 0.307 as $\delta$ exceeds $50$. In Fig. 8, we consider the scenario where there are $k$ connected suspect nodes. We observe that the correct detection probability significantly exceeds $1/k$ when $\delta>2$, and that reliable detection is achieved as $\delta$ grows large. In Fig. 9, we consider the scenario where there are two suspect nodes with a shortest path distance $d$. We observe that the correct detection probability significantly exceeds the prior $1/2$ when $\delta>2$. Furthermore, the detection probability increases as $d$ increases, and reliable detection is achieved when either $\delta$ or $d$ is sufficiently large.

In addition, to verify that even when $k$ becomes large our results in Theorem 3 will not degenerate to the case with no {\it a priori} knowledge, we let the suspect set cardinality $k$ range from 2 to 4000 in the scenario where the $k$ suspect nodes form a connected subgraph, and let $n = 1000$ nodes be eventually infected. From Fig. 10, we observe that the correct detection probability is always at least $1/2$, and accurately coincides with the theoretical prediction in Theorem 3.

\section{Conclusion}

In this paper, we have studied the problem of rooting out a single rumor source from a set of suspect nodes under an SI model, focusing on the performance analysis of the MAP source estimator. {In order to handle the presence of the suspect set, we have developed a key concept of {\it local rumor center} which greatly facilitates the analysis.} For regular tree-type networks, we have developed both finite and asymptotic detection performance results of the MAP estimator, using the P\'{o}lya's urn model in probability theory. We have investigated four representative scenarios, for each of which developing analytical results and shedding key insights into the behavior of the rumor spreading process in large-scale networks. The introduction of the {\it a priori} knowledge of suspect set dramatically improves the detection performance compared with the case of no {\it a priori} knowledge, and substantially enriches the scope of the rumor source estimation problem.

\appendices
\section*{Appendix}

\subsection{Proof of Lemma 6}

\begin{proof}[Proof of Lemma 6]
If $\max\{x_j,1 \leq j \leq m\}<n/2$, then from Proposition 1, we know that $s^\ast$ is the local rumor center w.r.t. the sub-neighborhood $N_l(s^\ast)=\{v_1,\ldots,v_m\}$ of $G_n$. Again from Proposition 1, we have $R(u,G_n) < R(s^\ast,G_n)$ for all $u \in T_{v_j}^{s^\ast}$ and $1 \leq j \leq m$. Therefore, we can make sure to correctly identify $s^\ast$ as the rumor source.

If $\max\{x_j,1 \leq j \leq m\}=n/2$, then from Proposition 1, we know that $s^\ast$ is the local rumor center w.r.t. the sub-neighborhood $N_l(s^\ast)=\{v_1,\ldots,v_m\}$ of $G_n$. Again from Proposition 1, there is only a node $u \in N_l(s^\ast)$ such that $R(u,G_n) = R(s^\ast,G_n)$, and $R(v,G_n) < R(s^\ast,G_n)$ for all $v \in \{T_{v_j}^{s^\ast},1 \leq j \leq m\} \setminus \{u\} \}$. Therefore, the probability to correctly identify $s^\ast$ as the rumor source is 1/2.

If $\max\{x_j,1 \leq j \leq m\}>n/2$, then from Proposition 1, we know that $s^\ast$ is not the local rumor center w.r.t. the sub-neighborhood $N_l(s^\ast)=\{v_1,\ldots,v_m\}$ of $G_n$. Therefore, we cannot identify $s^\ast$ as the rumor source.
\end{proof}

\subsection{Proof of \eqref{eq28}}

Here, we only present the detailed proof of \eqref{eq28} when $n$ is even. The case when $n$ is odd can be deduced similarly.

When $n$ is even, we have
\begin{eqnarray*}
\label{eq101}
\mathbf{P_c}(n)&=&\frac{1}{2}\sum_{\max\{x_1,x_2\}=n/2}{\mathbf{P_G}}\bigg[\bigcap_{j=1}^{2}(X_j=x_j)\bigg]\\
&&+\sum_{\max\{x_1,x_2\}<n/2}{\mathbf{P_G}}\bigg[\bigcap_{j=1}^{2}(X_j=x_j)\bigg]\\
&=&\frac{1}{2} \mathbf{P_G}\bigg[\frac{n-2}{2},\frac{n}{2}\bigg]+\frac{1}{2} \mathbf{P_G}\bigg[\frac{n}{2},\frac{n-2}{2}\bigg],
\end{eqnarray*}
where we use the short form of ${\mathbf{P_G}}\bigg[\bigcap_{j=1}^{2}(X_j=x_j)\bigg]$, just as ${\mathbf{P_G}}\bigg[x_j,1 \leq j \leq 2\bigg]$.

From \eqref{eq27}, we have

\small
\begin{eqnarray*}
\label{eq102}
\mathbf{P_c}(n)&=&\frac{1}{2^n} \frac{(n-1)!}{((n-2)/2)! (n/2)!}
+\frac{1}{2^n} \frac{(n-1)!}{(n/2)! ((n-2)/2)!}\\
&=&\frac{1}{2^{n-1}} \binom{n-1}{(n-2)/2}.
\end{eqnarray*}
\normalsize

\subsection{Proof of \eqref{eq31}}

Here, we only present the detailed proof of \eqref{eq31} when $n$ is odd. The case when $n$ is even can be deduced similarly.

When $n$ is odd, we have

\small
\begin{eqnarray*}
\label{eq105}
\mathbf{P_c}(n)&=&\frac{1}{2}\sum_{\max\{x_j,1 \leq j \leq 3\}=n/2}{\mathbf{P_G}}\bigg[\bigcap_{j=1}^{3}(X_j=x_j)\bigg]\\
&&+\sum_{\max\{x_j,1 \leq j \leq 3\}<n/2}{\mathbf{P_G}}\bigg[\bigcap_{j=1}^{3}(X_j=x_j)\bigg]\\
&=&\mathbf{P_G}\bigg[0,\frac{n-1}{2},\frac{n-1}{2}\bigg]\\
&&+\mathbf{P_G}\bigg[1,\frac{n-1}{2},\frac{n-3}{2}\bigg] + \mathbf{P_G}\bigg[1,\frac{n-3}{2},\frac{n-1}{2}\bigg]\\
&&\vdots\\
&&+\mathbf{P_G}\bigg[\frac{n-1}{2},\frac{n-1}{2},0\bigg] +\cdots+ \mathbf{P_G}\bigg[\frac{n-1}{2},0,\frac{n-1}{2}\bigg],
\end{eqnarray*}

\normalsize
\noindent where we use the short form of ${\mathbf{P_G}}\bigg[\bigcap_{j=1}^{3}(X_j=x_j)\bigg]$, just as ${\mathbf{P_G}}\bigg[x_j,1 \leq j \leq 3\bigg]$.

From \eqref{eq30}, we have
\begin{equation*}
\label{eq106}
\mathbf{P_c}(n) = \frac{2}{n(n+1)} \sum_{x=0}^{x=(n-1)/2}{(x+1)} = \frac{1}{4}+\frac{3}{4n}.
\end{equation*}

\subsection{Proof of Lemma 7}

\begin{proof}[Proof of Lemma 7]
Since the source $s^\ast$ has $m$ ($1 \leq m \leq \delta$) neighbors in the suspect set $S$, from Proposition 1 and Lemma 6, we have
\begin{eqnarray*}
\label{eq109}
\mathbf{P_c}(n|s^\ast) &\geq& \mathbf{P_G}\bigg[\bigcap_{j=1}^{m}E_j\bigg]=1-\mathbf{P_G}\bigg[\bigcup_{j=1}^{m}E_j^c\bigg]\\
&\stackrel{(a)}{\geq}& 1-\sum_{j=1}^{m}\mathbf{P_G}\bigg[E_j^c\bigg] \stackrel{(b)}{=} 1-m\mathbf{P_G}\bigg[E_1^c\bigg].
\end{eqnarray*}
Above, (a) is by the union bound over events $E_1^c, \ldots, E_m^c$, and (b) by symmetry.

Again using Proposition 1 and Lemma 6, we have
\begin{eqnarray*}
\label{eq110}
\mathbf{P_c}(n|s^\ast) &\leq& \mathbf{P_G}\bigg[\bigcap_{i=1}^{m}F_i\bigg]=1-\mathbf{P_G}\bigg[\bigcup_{i=1}^{m}F_i^c\bigg]\\
&\stackrel{(a)}{=}& 1-\sum_{i=1}^{m}\mathbf{P_G}\bigg[F_i^c\bigg] \stackrel{(b)}{=} 1-m\mathbf{P_G}\bigg[F_1^c\bigg].
\end{eqnarray*}
Above, (a) follows from the fact that events $F_1^c,\ldots,F_m^c$ are disjoint since there is at most a subtree such that the number of nodes in it is more than $n/2$, and (b) from symmetry.
\end{proof}

\subsection{Proof of \eqref{eq43}}

Here, we only present the detailed proof of \eqref{eq43} when $m=1$ with even $n$. The case when $m=1$ with odd $n$ can be deduced similarly, and the case when $m=2$ is the same as that in Appendix-B.

For $m=1$ and $n$ is even, we have

\small
\begin{eqnarray*}
\label{eq112}
\mathbf{P_c}(n|s^\ast)&=&\frac{1}{2}\sum_{x_1=n/2}{\mathbf{P_G}}\bigg[\bigcap_{j=1}^{2}(X_j=x_j)\bigg]
+\sum_{x_1<n/2}{\mathbf{P_G}}\bigg[\bigcap_{j=1}^{2}(X_j=x_j)\bigg]\\
&=&\frac{1}{2} \mathbf{P_G}\bigg[\frac{n}{2},\frac{n-2}{2}\bigg]
+ \mathbf{P_G}\bigg[\frac{n-2}{2},\frac{n}{2}\bigg] +\cdots+ \mathbf{P_G}\bigg[0,n-1\bigg],
\end{eqnarray*}

\normalsize
\noindent where we use the short form of ${\mathbf{P_G}}\bigg[\bigcap_{j=1}^{2}(X_j=x_j)\bigg]$, just as ${\mathbf{P_G}}\bigg[x_j,1 \leq j \leq 2\bigg]$.

From \eqref{eq42}, we have
\begin{eqnarray*}
\label{eq113}
\mathbf{P_c}(n|s^\ast)&=&\frac{1}{2^n} \binom{n-1}{n/{2}} + \frac{1}{2^{n-1}} \sum_{x=0}^{(n-2)/{2}}{\binom{n-1}{x}}\\
&=&\frac{1}{2} + \frac{1}{2^n} \binom{n-1}{(n-2)/2}.
\end{eqnarray*}

\subsection{Proof of \eqref{eq47}}

Here, we only present the detailed proof of \eqref{eq47} when $m=2$ with odd $n$. The case when $m=3$ is the same as that in Appendix-C, and the other cases can be deduced similarly.

For $m=2$ and $n$ is odd, we have

\small
\begin{eqnarray*}
\label{eq116}
\mathbf{P_c}(n|s^\ast)&=&\frac{1}{2}\sum_{\max\{x_j,1 \leq j \leq 2\}=n/2}{\mathbf{P_G}}\bigg[\bigcap_{j=1}^{3}(X_j=x_j)\bigg]\\
&&+\sum_{\max\{x_j,1 \leq j \leq 2\}<n/2}{\mathbf{P_G}}\bigg[\bigcap_{j=1}^{3}(X_j=x_j)\bigg]\\
&=&\mathbf{P_G}\bigg[0,\frac{n-1}{2},\frac{n-1}{2}\bigg] +\cdots+ \mathbf{P_G}\bigg[0,0,n-1\bigg]\\
&&+\mathbf{P_G}\bigg[1,\frac{n-1}{2},\frac{n-3}{2}\bigg] +\cdots+ \mathbf{P_G}\bigg[1,0,n-2\bigg]\\
&&\vdots\\
&&+\mathbf{P_G}\bigg[\frac{n-1}{2},\frac{n-1}{2},0\bigg] +\cdots+ \mathbf{P_G}\bigg[\frac{n-1}{2},0,\frac{n-1}{2}\bigg],
\end{eqnarray*}

\normalsize
\noindent where we use the short form of ${\mathbf{P_G}}\bigg[\bigcap_{j=1}^{3}(X_j=x_j)\bigg]$, just as ${\mathbf{P_G}}\bigg[x_j,1 \leq j \leq 3\bigg]$.

From \eqref{eq46}, we have
\begin{equation*}
\label{eq117}
\mathbf{P_c}(n|s^\ast) = \frac{2}{n(n+1)} \sum_{x=0}^{x=(n-1)/2}{\frac{n+1}{2}} = \frac{1}{2} + \frac{1}{2n}.
\end{equation*}

\subsection{Proof of \eqref{eq64}}

Here, we only present the detailed proof of \eqref{eq64} when $n$ is odd with even $d$. The other three cases can be deduced similarly.

Since $z_1 \geq (n+d+1)/2$ if and only if $R(s^\ast,G_n) \leq R(s_2,G_n)$, thus from \eqref{eq63}, we have

\small
\begin{eqnarray*}
\label{eq120}
\mathbf{P_e}(n)&=&\frac{1}{2}\sum_{z_1=(n+d+1)/2}{\mathbf{P_G}}\left[Z_1=z_1\right]
+\sum_{z_1>(n+d+1)/2}{\mathbf{P_G}}\left[Z_1=z_1\right]\\
&=&\frac{1}{2^n}\binom{n-1}{(n+d+1)/2}
+\frac{1}{2^{n-1}}\sum_{z_1=(n+d+3)/2}^{n-1}{\binom{n-1}{z_1}}\\
&=&\frac{1}{2}-\frac{1}{2^n}\sum_{z_1=(n-d-1)/2}^{(n+d+1)/2}{\binom{n-1}{z_1}}.
\end{eqnarray*}

\subsection{Proof of \eqref{eq67}}

Here, we only present the detailed proof of \eqref{eq67} when $n$ is even. The case when $n$ is odd can be deduced similarly.

Since $z_1+z_2 \geq n$ if and only if $R(s^\ast,G_n) \leq R(s_2,G_n)$. Besides, $z_1 \geq z_2+1$, thus $z_1 \geq (n+1)/2$. Therefore, from \eqref{eq66}, we have

\small
\begin{eqnarray*}
\label{eq121}
\mathbf{P_e}(n)&=&\frac{1}{2}\sum_{z_1+z_2=n}{\mathbf{P_G}}\left[\bigcap_{h=1}^{2}(Z_h=z_h)\right]
+\sum_{z_1+z_2>n}{\mathbf{P_G}}\left[\bigcap_{h=1}^{2}(Z_h=z_h)\right]\\
&=&\frac{1}{2}{\mathbf{P_G}}\bigg[\frac{n}{2}+1,\frac{n}{2}-1\bigg] +\cdots+ \frac{1}{2}{\mathbf{P_G}}\bigg[n-1,1\bigg]\\
&+&{\mathbf{P_G}}\bigg[\frac{n}{2}+1,\frac{n}{2}\bigg]\\
&+&{\mathbf{P_G}}\bigg[\frac{n}{2}+2,\frac{n}{2}-1\bigg] +\cdots+ {\mathbf{P_G}}\bigg[\frac{n}{2}+2,\frac{n}{2}+1\bigg]\\
&\vdots&\\
&+&{\mathbf{P_G}}\bigg[n-1,2\bigg] +\cdots+ {\mathbf{P_G}}\bigg[n-1,n-2\bigg]\\
&=&\frac{2}{n(n+1)}\frac{n/2-1}{n/2+1}\left(\frac{1}{2}+1\right)+\frac{2}{n(n+1)}\frac{n/2-2}{n/2+2}\left(\frac{1}{2}+3\right)\\
&&+\cdots+\frac{2}{n(n+1)}\frac{n/2-(n-2)/2}{n/2+(n-2)/2}\left(\frac{1}{2}+n-3\right),
\end{eqnarray*}

\normalsize
\noindent where we use the short form of ${\mathbf{P_G}}\bigg[\bigcap_{h=1}^{3}(Z_h=z_h)\bigg]$, just as ${\mathbf{P_G}}\bigg[z_h,1 \leq h \leq 3\bigg]$. As $n \rightarrow \infty$, we have
\begin{equation*}
\label{eq122}
\lim_{n \rightarrow \infty}\mathbf{P_e}(n)\approx0.114.
\end{equation*}

\subsection{Proof of Lemma 8}

\begin{proof}[Proof of Lemma 8]
Consider the case when the rumor source $s^\ast$ has $m$ ($1 \leq m \leq \delta$) suspect neighbors $N_l(s^\ast)=\{v_1,\ldots,v_m\}$. From Proposition 1, if $s^\ast$ is the local rumor center w.r.t. $N_l(s^\ast)$ of $G_n$, then $R(s^\ast,G_n) \geq R(u,G_n)$ for all $u \in S \bigcap N_l(s^\ast)$, and $R(s^\ast,G_n) > R(u^\prime,G_n)$ for all $u \in S \setminus \{N_l(s^\ast) \bigcup s^\ast\}$.

When all the $m$ suspect neighbors of $s^\ast$ are neighboring suspect nodes of $s^\ast$, from Proposition 1, the MAP estimator can correctly identify $s^\ast$ as the estimate $\hat{s}$ if and only if $s^\ast$ is the local rumor center. However, in general cases, the MAP estimator can correctly identify $s^\ast$ as the estimate $\hat{s}$ even if $s^\ast$ is not a local rumor center.
\end{proof}

\subsection{Proof of Lemma 9}

\begin{proof}[Proof of Lemma 9]
Define $E_j^{=}=\{X_j = {n}/{2}\}$ and $E_j^{>}=\{X_j > {n}/{2}\}$, $1 \leq j \leq m$. Note that $\{E_j^{=},E_j^{>},1 \leq j \leq m\}$ are all disjoint. For the error detection probability $\mathbf{P_e}(n|s^\ast)=1-\mathbf{P_c}(n|s^\ast)$, from Proposition 1 and Lemma 6, we have
\begin{eqnarray*}
\label{eq124}
\mathbf{P_e}(n|s^\ast) &=& 0.5\mathbf{P_G}\bigg[\bigcup_{j=1}^{m} E_j^{=} \bigg] + \mathbf{P_G}\bigg[\bigcup_{j=1}^{m} E_j^{>} \bigg]\\
&=& m\left(0.5{\mathbf{P_G}}\left(X_1=\frac{n}{2}\right)+\sum_{x_1>n/2}{\mathbf{P_G}}\left(X_1=x_1\right)\right),
\end{eqnarray*}
where the last equality follows from symmetry.
\end{proof}

\subsection{Proof of Corollary 10}

For Corollary 10-{\romannumeral1} with fixed node degree $\delta \geq 2$, from \eqref{eq88}, we should prove that $\mathbf{P_{e1}}(n)=\left(1-\mathbf{P_c}(n)\right)/\delta=0.5{\mathbf{P_G}}\left(X_1=\frac{n}{2}\right)+\sum_{x_1>n/2}{\mathbf{P_G}}\left(X_1=x_1\right)$ increases with $n$ ($n \geq 2$). Note that $\mathbf{P_{e1}}(n)$ is the error detection probability of the MAP estimator to identify the neighboring suspect node $s^\ast_1$ of the rumor source $s^\ast$ as the estimate $\hat{s}$. In the following, we deduce the relation of $\mathbf{P_{e1}}(n)$ between $n=2i$ and $n=2i+1$, and that between $n=2i+1$ and $n=2i+2$, for all $i \in N^+$.

\begin{proof}[Proof of Corollary 10-{\romannumeral1}]
First, consider the relation of $\mathbf{P_{e1}}(n)$ between $n=2i$ and $n=2i+1$. From \eqref{eq88}, we have
\begin{equation*}
\label{eq130}
\mathbf{P_{e1}}(n=2i+1)=\sum_{x_1=i+1}^{2i}{\mathbf{P_G}}\left(X_1=x_1 | n=2i+1\right),
\end{equation*}
and

\small
\begin{equation*}
\label{eq131}
\mathbf{P_{e1}}(n=2i)=0.5{\mathbf{P_G}}\left(X_1=i | n=2i\right)+\sum_{x_1=i+1}^{2i-1}{\mathbf{P_G}}\left(X_1=x_1 | n=2i\right).
\end{equation*}

\normalsize
\noindent where $\mathbf{P_G}\left(X_1=x_1 | n\right)$ is given by \eqref{eq240} conditioned on $n$.

When $x_1=2i$, from \eqref{eq240}, we have

\small
\begin{eqnarray*}
\label{eq132}
&&\mathbf{P_G}\left(X_1=2i | n=2i+1\right)\\
&=&
\binom{2i}{2i}\frac{\prod_{j=1}^{2}b_j(b_j+\epsilon)\cdots(b_j+(x_j-1)\epsilon)}{\delta(\delta+\epsilon)\cdots(\delta+(2i-1)\epsilon)}\\
&=&
\frac{1+(2i-1)(\delta-2)}{\delta+(2i-1)(\delta-2)}\binom{2i-1}{2i-1}\frac{\prod_{j=1}^{2}b_j(b_j+\epsilon)\cdots(b_j+(x_j-1)\epsilon)}{\delta(\delta+\epsilon)\cdots(\delta+(2i-2)\epsilon)}\\
&=&
\frac{1+(2i-1)(\delta-2)}{2+2i(\delta-2)}\mathbf{P_G}\left(X_1=2i-1 | n=2i\right),
\end{eqnarray*}

\normalsize
\noindent where $b_1=1$, $b_2=\delta-1$, $\epsilon=\delta-2$ and $x_2=n-x_1-1$.

When $x_1=i+1, \ldots, 2i-1$, from \eqref{eq240}, we have

\small
\begin{eqnarray*}
\label{eq133}
&&\mathbf{P_G}\left(X_1=x+1 | n=2i+1\right)\\
&=&
\binom{2i}{x+1}\frac{\prod_{j=1}^{2}b_j(b_j+\epsilon)\cdots(b_j+(x_j-1)\epsilon)}{\delta(\delta+\epsilon)\cdots(\delta+(2i-1)\epsilon)}\\
&=&
\frac{2i}{x+1}\frac{1+x(\delta-2)}{\delta+(2i-1)(\delta-2)}\binom{2i-1}{x}\frac{\prod_{j=1}^{2}b_j\cdots(b_j+(x_j-1)\epsilon)}{\delta\cdots(\delta+(2i-2)\epsilon)}\\
&=&
\frac{2i}{x+1}\frac{1+x(\delta-2)}{\delta+(2i-1)(\delta-2)}\mathbf{P_G}\left(X_1=x | n=2i\right)\\
&=&
\frac{1+x(\delta-2)}{\delta+(2i-1)(\delta-2)}\mathbf{P_G}\left(X_1=x | n=2i\right)\\
&&+
\frac{2i-x-1}{x+1}\frac{1+x(\delta-2)}{\delta+(2i-1)(\delta-2)}\mathbf{P_G}\left(X_1=x | n=2i\right)\\
&=&
\frac{1+x(\delta-2)}{\delta+(2i-1)(\delta-2)}\mathbf{P_G}\left(X_1=x | n=2i\right)\\
&&+
\frac{\delta-1+(2i-x-2)(\delta-2)}{\delta+(2i-1)(\delta-2)}\binom{2i-1}{x+1}\frac{\prod_{j=1}^{2}b_j\cdots(b_j+(x_j-1)\epsilon)}{\delta\cdots(\delta+(2i-2)\epsilon)}\\
&=&
\frac{1+x(\delta-2)}{2+2i(\delta-2)}\mathbf{P_G}\left(X_1=x | n=2i\right)\\
&&+
\frac{1+(2i-x-1)(\delta-2)}{2+2i(\delta-2)}\mathbf{P_G}\left(X_1=x+1 | n=2i\right),
\end{eqnarray*}

\normalsize
\noindent where $b_1=1$, $b_2=\delta-1$, $\epsilon=\delta-2$ and $x_2=n-x_1-1$.

Using the above relations and merging the adjacent items, we have
\begin{eqnarray*}
\label{eq134}
&&\mathbf{P_{e1}}(n=2i+1)=\sum_{x_1=i+1}^{2i}{\mathbf{P_G}}\left(X_1=x_1 | n=2i+1\right)\\
&=&
\frac{1+(2i-1)(\delta-2)}{2+2i(\delta-2)}\mathbf{P_G}\left(X_1=2i-1 | n=2i\right)\\
&&+
\frac{1+1(\delta-2)}{2+2i(\delta-2)}\mathbf{P_G}\left(X_1=2i-1 | n=2i\right)\\
&&+
\frac{1+(2i-2)(\delta-2)}{2+2i(\delta-2)}\mathbf{P_G}\left(X_1=2i-2 | n=2i\right)\\
&&\vdots\\
&&+
\frac{1+i(\delta-2)}{2+2i(\delta-2)}\mathbf{P_G}\left(X_1=i | n=2i\right)\\
&=&
0.5{\mathbf{P_G}}\left(X_1=i | n=2i\right)+\sum_{x_1=i+1}^{2i-1}{\mathbf{P_G}}\left(X_1=x_1 | n=2i\right)\\
&=&\mathbf{P_{e1}}(n=2i).
\end{eqnarray*}

Now, consider the relation of $\mathbf{P_{e1}}(n)$ between $n=2i+1$ and $n=2i+2$. From \eqref{eq88}, we have
\begin{eqnarray*}
\label{eq135}
\mathbf{P_{e1}}(n=2i+2)&=&0.5{\mathbf{P_G}}\left(X_1=i+1 | n=2i+2\right)\\
&&+
\sum_{x_1=i+2}^{2i+1}{\mathbf{P_G}}\left(X_1=x_1 | n=2i+2\right),
\end{eqnarray*}
and
\begin{equation*}
\label{eq136}
\mathbf{P_{e1}}(n=2i+1)=\sum_{x_1=i+1}^{2i}{\mathbf{P_G}}\left(X_1=x_1 | n=2i+1\right).
\end{equation*}
where $\mathbf{P_G}\left(X_1=x_1 | n\right)$ is given by \eqref{eq240} conditioned on $n$.

When $x_1=2i+1$, similarly from \eqref{eq240}, we have
\begin{eqnarray*}
\label{eq137}
&&\mathbf{P_G}\left(X_1=2i+1 | n=2i+2\right)\\
&=&
\frac{1+2i(\delta-2)}{2+(2i+1)(\delta-2)}\mathbf{P_G}\left(X_1=2i | n=2i+1\right).
\end{eqnarray*}

When $x_1=i+1, \ldots, 2i$, similarly from \eqref{eq240}, we have
\begin{eqnarray*}
\label{eq138}
&&\mathbf{P_G}\left(X_1=x+1 | n=2i+2\right)\\
&=&
\frac{1+x(\delta-2)}{2+(2i+1)(\delta-2)}\mathbf{P_G}\left(X_1=x | n=2i+1\right)\\
&&+
\frac{1+(2i-x)(\delta-2)}{2+(2i+1)(\delta-2)}\mathbf{P_G}\left(X_1=x+1 | n=2i+1\right).
\end{eqnarray*}

Using the above relations and merging the adjacent items, we have

\small
\begin{eqnarray*}
\label{eq139}
&&\mathbf{P_{e1}}(n=2i+2)\\
&=&
0.5{\mathbf{P_G}}\left(X_1=i+1 | n=2i+1\right)+\sum_{x_1=i+2}^{2i+1}{\mathbf{P_G}}\left(X_1=x_1 | n=2i+2\right)\\
&=&
\frac{1+2i(\delta-2)}{2+(2i+1)(\delta-2)}\mathbf{P_G}\left(X_1=2i | n=2i+1\right)\\
&&+
\frac{1+1(\delta-2)}{2+(2i+1)(\delta-2)}\mathbf{P_G}\left(X_1=2i | n=2i+1\right)\\
&&+
\frac{1+(2i-1)(\delta-2)}{2+(2i+1)(\delta-2)}\mathbf{P_G}\left(X_1=2i-1 | n=2i+1\right)\\
&&\vdots\\
&&+
\frac{1+(i+1)(\delta-2)}{2+(2i+1)(\delta-2)}\mathbf{P_G}\left(X_1=i+1 | n=2i+1\right)\\
&&+
0.5\frac{1+i(\delta-2)}{2+(2i+1)(\delta-2)}\mathbf{P_G}\left(X_1=i+1 | n=2i+1\right)\\
&&+
0.5\frac{1+i(\delta-2)}{2+(2i+1)(\delta-2)}\mathbf{P_G}\left(X_1=i | n=2i+1\right)\\
&=&
\sum_{x_1=i+1}^{2i}{\mathbf{P_G}}\left(X_1=x_1 | n=2i+1\right)\\
&&+
\frac{1}{2i}\frac{1+i(\delta-2)}{2+(2i+1)(\delta-2)}\mathbf{P_G}\left(X_1=i+1 | n=2i+1\right)\\
&=&\mathbf{P_{e1}}(n=2i+1)\\
&&+
\frac{1}{2i}\frac{1+i(\delta-2)}{2+(2i+1)(\delta-2)}\mathbf{P_G}\left(X_1=i+1 | n=2i+1\right).
\end{eqnarray*}

\normalsize
In summary, we have
\begin{equation*}
\label{eq140}
\mathbf{P_{e1}}(n=2i+1)=\mathbf{P_{e1}}(n=2i),
\end{equation*}
and
\begin{eqnarray*}
\label{eq141}
\mathbf{P_{e1}}(n=2i+2)=\mathbf{P_{e1}}(n=2i+1)\quad\quad\quad\quad\quad\quad\quad\quad\quad\quad\\
+\frac{1}{2i}\frac{1+i(\delta-2)}{2+(2i+1)(\delta-2)}\mathbf{P_G}\left(X_1=i+1 | n=2i+1\right),
\end{eqnarray*}
for all $i \in N^+$. Therefore, the proof of Corollary 10-{\romannumeral1} is completed.
\end{proof}

For Corollary 10-{\romannumeral2} with fixed infection number $n \geq 2$, from \eqref{eq88}, we should prove that $\delta \cdot \mathbf{P_{e1}}(n)=1-\mathbf{P_c}(n)=\delta\left(0.5{\mathbf{P_G}}\left(X_1=\frac{n}{2}\right)+\sum_{x_1>n/2}{\mathbf{P_G}}\left(X_1=x_1\right)\right)$ decreases with $\delta$ ($\delta \geq 2$). In the following, we show that the derivative of $\delta \cdot \mathbf{P_{e1}}(n)$ w.r.t. $\delta$ is non-positive using the mathematical induction.

\begin{proof}[Proof of Corollary 10-{\romannumeral2}]
First, consider $n=2$, and $\delta \cdot \mathbf{P_{e1}}(n)=1/2$. Therefore, the derivative of $\delta \cdot \mathbf{P_{e1}}(n=2)$ w.r.t. $\delta$ is non-positive.

Then, suppose that the derivative of $\delta \cdot \mathbf{P_{e1}}(n=2i)$ w.r.t. $\delta$ is non-positive, where $i \in N^+$. Since $\mathbf{P_{e1}}(n=2i+1)=\mathbf{P_{e1}}(n=2i)$, thus the derivative of $\delta \cdot \mathbf{P_{e1}}(n=2i+1)$ w.r.t. $\delta$ is non-positive. Next, we prove that the derivative of $\delta \cdot \mathbf{P_{e1}}(n=2i+2)$ w.r.t. $\delta$ is non-positive.

Since we have
\begin{eqnarray*}
\label{eq142}
\mathbf{P_{e1}}(n=2i+2)=\mathbf{P_{e1}}(n=2i+1)\quad\quad\quad\quad\quad\quad\quad\quad\quad\\
+\frac{1}{2i}\frac{1+i(\delta-2)}{2+(2i+1)(\delta-2)}\mathbf{P_G}\left(X_1=i+1 | n=2i+1\right),
\end{eqnarray*}
the proof will be completed if we show that the derivative of $\frac{1}{2i}\frac{1+i(\delta-2)}{2+(2i+1)(\delta-2)} \cdot \delta \cdot \mathbf{P_G}\left(X_1=i+1 | n=2i+1\right)$ w.r.t. $\delta$ is non-positive. In the following, we use the logarithmic function of $\frac{1}{2i}\frac{1+i(\delta-2)}{2+(2i+1)(\delta-2)} \cdot \delta \cdot \mathbf{P_G}\left(X_1=i+1 | n=2i+1\right)$ for convenience.

Letting $g_1(\delta | i)=\log\left(\frac{1}{2i}\frac{1+i(\delta-2)}{2+(2i+1)(\delta-2)}\right)$, then we have
\begin{eqnarray*}
\label{eq143}
g_1^\prime(\delta | i)&=&\frac{i}{1+i(\delta-2)}-\frac{2i+1}{2+(2i+1)(\delta-2)}\\
&=&\frac{-1}{(1+i(\delta-2))(2+(2i+1)(\delta-2))}.
\end{eqnarray*}

Let $g_2(\delta | i)=\log\left(\delta \cdot \mathbf{P_G}\left(X_1=i+1 | n=2i+1\right)\right)$, and we use the mathematical induction once again so as to prove that the derivative of $g_2(\delta | i)$ w.r.t. $\delta$ is non-positive for all fixed $i$.

Considering $i=1$, then we have $g_2(\delta | i=1)=\log\left(\frac{1}{2}\right)$. Therefore, the derivative of $g_2(\delta | i=1)$ w.r.t. $\delta$ is non-positive. Suppose that the derivative of $g_2(\delta | i)$ w.r.t. $\delta$ is non-positive, where $i \in N^+$. From \eqref{eq240}, we have
\begin{equation*}
\label{eq145}
g_2(\delta | i+1)=g_2(\delta | i)-\log(2)+\log\left(\frac{1+i(\delta-2)}{\delta+2i(\delta-2)}\right).
\end{equation*}
Taking its derivative, we have
\begin{eqnarray*}
g_2^\prime(\delta | i+1)&=&g_2^\prime(\delta | i)+\frac{i}{1+i(\delta-2)}-\frac{2i+1}{\delta+2i(\delta-2)}\\
&=&g_2^\prime(\delta | i)+\frac{-1}{(1+i(\delta-2))(2+(2i+1)(\delta-2))}.
\end{eqnarray*}
Therefore, the derivative of $g_2(\delta | i)$ w.r.t. $\delta$ is non-positive for all fixed $i$.

In summary, we have $g_1^\prime(\delta | i) \leq 0$ and $g_2^\prime(\delta | i) \leq 0$, and $\frac{1}{2i}\frac{1+i(\delta-2)}{2+(2i+1)(\delta-2)} \cdot \delta \cdot \mathbf{P_G}\left(X_1=i+1 | n=2i+1\right)=\exp{\left(g_1(\delta | i) \cdot g_2(\delta | i)\right)}$. Combing these results together, we have proved that the derivative of $\delta \cdot \mathbf{P_{e1}}(n=2i+2)$ w.r.t. $\delta$ is non-positive. Therefore, the proof of Corollary 10-{\romannumeral2} is completed.
\end{proof}



\bibliographystyle{IEEEtran}	
\bibliography{myrefs}           

\end{document}